\documentclass[superscriptaddress, amsmath,amssymb, pre, twocolumn,floatfix,showkeys]{revtex4-1}

\usepackage{graphicx}
\usepackage{dcolumn}
\usepackage{bm}
\usepackage{hyperref}
\usepackage{marginnote}
\usepackage{subfigure}

\usepackage{color}

\begin{document}

\title{Nonlinear mechanosensation in fiber networks}

\author{Estelle Berthier}
\affiliation{Arnold-Sommerfeld-Center for Theoretical Physics and Center for NanoScience, Ludwig-Maximilians-Universit\"at M\"unchen, D-80333 M\"unchen, Germany}
\author{Haiqian Yang}
\affiliation{Department of Mechanical Engineering, Massachusetts Institute of Technology, Cambridge, MA, USA}
\author{Ming Guo}
\affiliation{Department of Mechanical Engineering, Massachusetts Institute of Technology, Cambridge, MA, USA}
\author{Pierre Ronceray$^*$}
\affiliation{Aix Marseille Univ, CNRS, CINAM,Turing Center for Living Systems, Marseille, France}
\email{pierre.ronceray@univ-amu.fr}
\author{Chase P. Broedersz$^*$}
\affiliation{Arnold-Sommerfeld-Center for Theoretical Physics and Center for NanoScience, Ludwig-Maximilians-Universit\"at M\"unchen, D-80333 M\"unchen, Germany}
\affiliation{Vrije Universiteit Amsterdam,  Department of Physics and Astronomy, 1081 HV Amsterdam, The Netherlands}
\email{c.p.broedersz@vu.nl}

\begin{abstract}
In a diversity of physiological contexts, eukaryotic cells adhere to an extracellular matrix (ECM), a disordered network with complex nonlinear mechanics. Such cells can perform mechanosensation: using local force probing they can measure and respond to their substrate's mechanical properties. It remains unclear, however, how the mechanical complexity of the ECM at the cellular scale impacts mechanosensation.  Here, we investigate the physical limits of mechanosensation  imposed by the inherent structural disorder and nonlinear elastic response of the ECM. Using a theoretical framework for disordered fiber networks, we find that the extreme mechanical heterogeneity that cells can locally sense with small probing forces is strongly reduced with increasing force. Specifically, we predict that the accuracy of mechanosensation  dramatically improves with force, following  a universal power law insensitive to constitutive details, which we quantitatively confirm using microrheology experiments in collagen and fibrin gels. We provide conceptual insights into this behavior by introducing a general model for nonlinear mechanosensation, based on the idea of an emergent  nonlinear length-scale associated with fiber buckling. This force-dependent length-scale enhances the range over which local  mechanical measurements are performed, thereby  averaging the response of a disordered network over an enlarged region. We show with an example how a cell can use this nonlinear mechanosensation to infer the macroscopic mechanical properties of a disordered ECM using local measurements. Together, our results demonstrate that cells can take advantage of the inherent nonlinearity of fibrous networks to robustly sense, control, and respond to their mechanical environment.
\end{abstract}

\keywords{mechanosensing, biopolymer networks, nonlinearity, structural disorder} 

\maketitle

Cell behavior is steered by various cues from their extracellular environment.  
Such cues include chemical, electrical, and topographic signals that regulate key cell functions such as migration~\cite{shellard2020all} and thereby impact processes ranging from embryonic development~\cite{wozniak2009mechanotransduction,heisenberg2013forces,campas2014quantifying} and tissue maintenance to disease progression~\cite{janmey2011mechanisms,wirtz2011physics}. 
In particular, there is growing evidence for mechanosensation: cells sense and respond to the mechanical properties of their environment~\cite{discher2005tissue,janmey2011mechanisms,vogel2006local,doyle2016mechanosensing,minner2014polymer}. The stiffness of the  cell's substrate can guide developmental processes \emph{in vivo}, such as axonal growth~\cite{koser2016mechanosensing}.
\emph{In vitro} model systems further revealed that  cells mechanically probe their substrate  and subsequently modify behaviors such as differentiation~\cite{engler2006matrix}, gene expression~\cite{chowdhury2010material}, and motility~\cite{lo2000cell,shellard2021durotaxis,dietrich2018guiding}.
It remains unclear, however, what mechanical information cells can perceive inside the complex environments they encounter naturally~\cite{janmey2020stiffness,wen2013effects}.

\emph{In vivo}, many cell types mechanically interact with the extracellular matrix (ECM) by adhering to network fibers and exerting local forces. The polymerization and gelation processes through which these collageneous matrices form result in an inherently disordered fiber network with large structural variations at the cell scale~\cite{lang2013estimating,frantz2010extracellular}. Consequently, the mechanical properties cells can locally perceive depend strongly on network location~\cite{grill2021directed,beroz2017physical,proestaki2019modulus,jones2015micromechanics,hayn2020inhomogeneities,proestaki2021effect}. This implies that cells face a highly heterogeneous mechanical environment in which the cell-scale linear stiffness, measured locally at different locations in a single network, exhibits relative variations as large as  between the  macroscopic stiffness of tissues as distinct as  brain (1~kPa) and bone (100~kPa)~\cite{beroz2017physical}. Thus, even if cellular  mechanosensors are ideal and can perfectly measure the local linear mechanical response, mechanosensation  remains limited by matrix disorder. Importantly however, cells can exert forces of up to few nanonewtons~\cite{trichet2012evidence,freyman2002fibroblast,tan2003cells,legant2010measurement} to probe their environment, easily exceeding the linear response regime of the ECM~\cite{han2018cell,van2016strain,hall2016fibrous}. Indeed, collagen networks  exhibit a pronounced nonlinear response at relatively small stresses or deformations. It is unclear how such nonlinearities impact cellular mechanosensation inside the disordered ECM.

The macroscopic nonlinear behavior of disordered fiber networks is well characterized both theoretically~\cite{mackintosh1995elasticity,huisman2010semiflexible,broedersz2014modeling,storm2005nonlinear,onck2005alternative,stein2011micromechanics,ban2019strong,wang2014long} and experimentally~\cite{licup2015stress,munster2013strain,kim2014structural,gardel2004elastic,lin2010origins,piechocka2010structural,jaspers2014ultra}. Nonlinearities arise through a range of effects, including constituent nonlinearities such as fiber buckling~\cite{conti2009cross} and entropic stiffening~\cite{storm2005nonlinear,burkel2017mechanical,gardel2004elastic},  or network nonlinearities arising from their low connectivity~\cite{wyart2008elasticity,sharma2016strain, broedersz2012filament}.
By contrast, the  nonlinear mechanics of the network in response to local probes at scales relevant to cell sensing remains less well understood ~\cite{head2005mechanical,liang2016heterogeneous,proestaki2019modulus,grill2021directed,abhilash2014remodeling,wang2014long}.
A nonlinear region emerges in the vicinity of a local force~\cite{han2018cell,burkel2017mechanical}, where fiber buckling and alignment result in a stress decay that is slower than in linear elasticity, consistent with cell-generated displacements~\cite{hall2016fibrous}. Recent discrete and continuous theoretical approaches established the role of buckling in the formation of this region~\cite{rosakis2015model,notbohm2015microbuckling,ronceray2016fiber,xu2015nonlinearities}, determined its spatial range~\cite{ronceray2016fiber,xu2015nonlinearities,sander2013alignment}, and its  impact on stress transmission~\cite{ronceray2016fiber,ronceray2019stress}. To investigate the ability of cells to infer matrix mechanics using local force probes, we thus need to understand the interplay between these nonlinearities and the structural disorder of the network.

Here, we investigate nonlinear mechanosensation from the perspective of a cell as an ideal mechanosensor inside a disordered matrix. Using a network model, we study the relation between bulk and local network properties. We derive a simple relation describing how the fiber constituent stiffening and density control the macroscopic nonlinear  mechanics, which explains the unique density-independence of the nonlinear stiffness of collagen networks~\cite{licup2015stress}. Surprisingly however, we find that these macroscopic nonlinear properties are largely irrelevant microscopically.
At the scale of a cell, the local mechanical response is strongly modulated by variations in fiber density. Using both theory and experiments, we discover a generic power law decay of the variability of local stiffness measurements with force.  For a range of fiber constitutive nonlinearities, the local stiffness becomes increasingly insensitive to network disorder.
To provide conceptual insights we develop a model for this nonlinear mechanosensation. Our model shows how large probing forces applied by a mechanosensor induce fiber buckling over an extended range, thereby effectively enhancing the length-scale over which a mechanical measurement is averaged in a disordered network. Thus, we here find that elastic nonlinearities can be exploited by cells to overcome the inherent disorder of their environment such that local measurements can be employed to accurately infer the macroscopic mechanical properties of the ECM.

\section*{Model for disordered nonlinear fiber networks}\label{model}

To investigate the consequences of structural heterogeneity in the ECM on the local mechanical environment cells can perceive, we build on a broadly used minimal model for a disordered fibrous matrix~\cite{broedersz2011criticality,broedersz2014modeling}.
In this model (Materials and Methods),  we introduce structural disorder by randomly depleting bonds on a regular lattice. The lattice fibers are represented by these bonds that are present with a probability $p$, setting the  fiber density.
Fiber bonds resist both bending and longitudinal deformations. Here, we describe the longitudinal fiber response with a nonlinear force-extension constitutive law (CL) $\tau = f(\epsilon)$, with $\epsilon$ the fiber's deformation and $\tau$ its tension. This CL is chosen to be asymmetric in compression and tension. Indeed, fibers buckle and soften under compression ($\epsilon <0$) and  stiffen beyond a characteristic tension, with a power law  increase $k\propto \tau^x$ of their differential stiffness $k=d\tau/d\epsilon$, where the exponent $x$ characterizes the stiffening mechanism.

\begin{figure}[t]
\centering
\includegraphics[width=\linewidth]{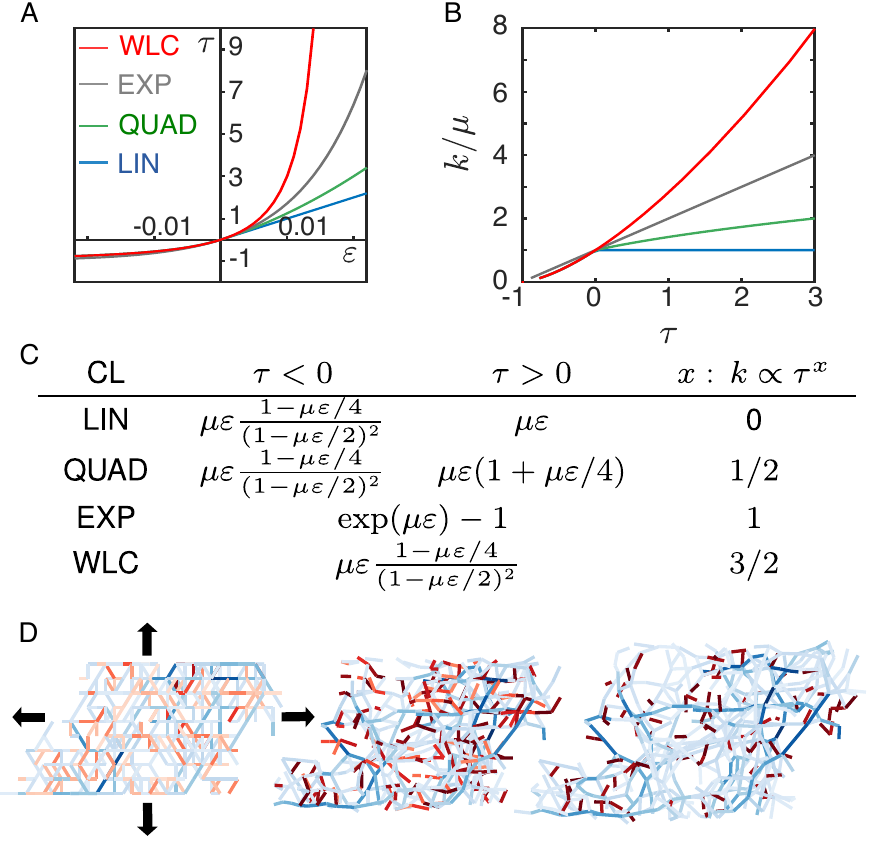}
\caption{\textbf{Nonlinear fiber network model.} Constitutive law (CL) of the fibers, A: bond tension ($\tau$) versus deformation ($\epsilon$), B: corresponding bond differential stiffness ($k$), normalized by the linear modulus ($\mu$) as a function of its tension, C: mathematical expression of CLs. D: Macroscopic loading of a 3D EXP-fiber network with $p=0.35$ at three dilatation strains $\gamma = 10^{-5}$, $8\cdot 10^{-2}$ and $2\cdot 10^{-1}$. Color code: low to high tensile  forces (resp. compressive forces) from light to dark blue (resp. red).}
\label{fig:model}
\end{figure}

Of particular interest is the case $x=1$, corresponding to an exponential CL (EXP, Figs.~\ref{fig:model}A-C),  
reflecting the empirically established stress-strain relationship of tendon and reconstituted collagen networks~\cite{fung1967elasticity,licup2015stress}.  To assess the effects of the nonlinear fiber micromechanics on cellular mechanosensing, we  also consider three other CLs, Figs.~\ref{fig:model}A-C. 
These CLs exhibit buckling-induced fiber softening and several distinct tensile responses described by a power law $k\propto \tau^x$: $x=0$ (LIN) corresponds to linear non-stiffening springs, $x=1/2$ (QUAD) to a quadratic force-extension relation, and $x=3/2$ (WLC) describes a divergent entropic force-extension relation of the worm-like chain model~\cite{mackintosh1995elasticity, gardel2004elastic}. Throughout this article, we use the mechanical equilibrium response to global and local loading of these model networks to study the ability of cells to glean information on the stiffness of their surrounding heterogeneous environment.

\section*{Density-independent nonlinear modulus of collagen}\label{macro}

\begin{figure*}[t]
  \centering
  \includegraphics[width = \textwidth ]{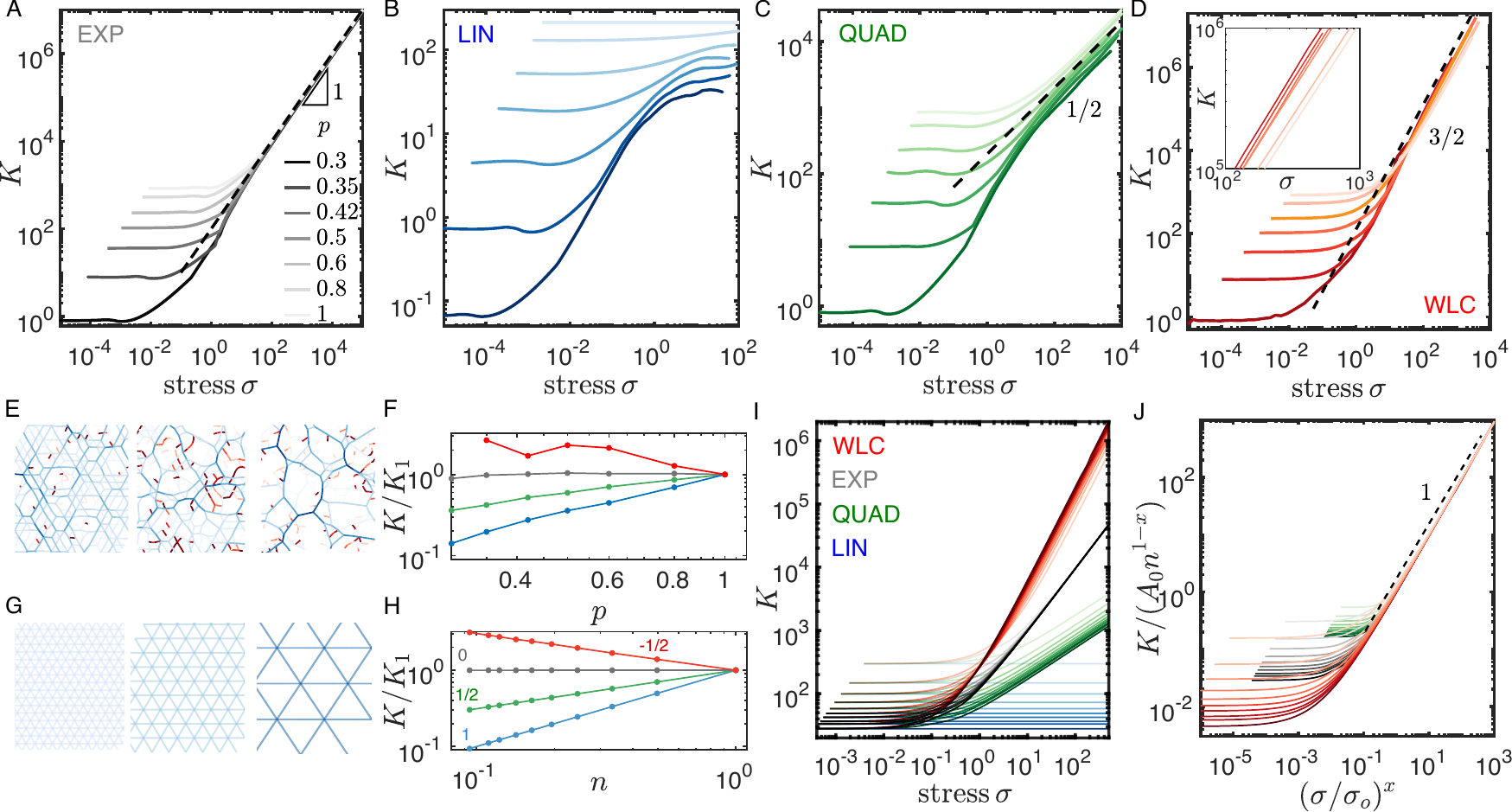}
  \caption{ \textbf{Stress- and fiber density-dependence of the bulk modulus.} 
  Macroscopic differential elastic modulus ($K$) versus bulk stress ($\sigma$) of 3D depleted networks with various fiber densities set by the bond occupation parameter $p$ (low to high value indicated by a dark to bright color)  and constituted of A: EXP-fibers, B: LIN-fibers, C: QUAD-fibers and D: WLC-fibers, the inset highlights the $p-$dependence in the large stress regime.
  Tension distribution in 2D EXP-fiber networks (same color code as Fig.~\ref{fig:model}C) of E: randomly depleted networks with, from left to right, $p= 0.5$, $0.6$ and $0.8$ at a fixed stress in the asymptotic large stress regime.
  F:  $p$-dependence of the large-stress modulus of 3D depleted networks. The normalization constant $K_1$ is the modulus for the $p=1$ homogeneous network.
  G: same as E for regular networks of decreasing fiber density. 
  H:  $n$-dependence of the large-stress modulus of the regular networks. The normalization constant $K_1$ is the modulus for the $n=1$ denser network.
  I: $K$ versus $\sigma$ of regular networks with low to high fiber densities for the different CLs, J: corresponding rescaled response, with $A_0$ and $\sigma_0$ CL-dependent constants.  }
  \label{fig:macro}
\end{figure*}

Cells can only locally probe the network to infer the mechanics of their surroundings. When using small probing forces, these local measurements are highly sensitive to density heterogeneity: the random architecture leads to some regions of the network being denser than others, thereby modulating the local mechanical properties~\cite{grill2021directed,beroz2017physical,proestaki2019modulus,jones2015micromechanics,hayn2020inhomogeneities,proestaki2021effect}. To understand how, we first investigate how fiber density affects the bulk stiffness of the network. While this dependency is well-understood in the linear, low-stress regime~\cite{das2012redundancy,broedersz2011criticality}, this is not the case in the nonlinear stiffening regime arising at larger stress, which is relevant for cell-ECM mechanical interactions~\cite{han2018cell,van2016strain,hall2016fibrous}.

We simulate the  response to a dilatation strain of  networks with varying fiber density  of EXP-fibers representing collagen (Fig.~\ref{fig:model}D, Fig.~\ref{fig:macro}A). The network stiffness is quantified by the
differential bulk modulus $K=d\sigma/d\gamma$, with $\sigma$ and $\gamma$ the macroscopic stress and strain (Materials and Methods). 
At low stress, the response is linear: network stiffness is stress-independent and, intuitively, increases with fiber density. In contrast, after a cross-over at intermediate stresses, the modulus increases with stress as a power law $K\propto \sigma$ that reflects the fiber constitutive nonlinearity. Strikingly, in this regime the macroscopic elastic responses \textit{converge} to a stress-controlled value, insensitive to fiber density, consistent with  macrorheology experiments on collagen gels~\cite{licup2015stress}.

To elucidate the stress- and density-dependence of the nonlinear macroscopic response of collagen, we propose a  differentially affine model. Indeed, we observe that at  large stresses a tense subnetwork that carries most of the stress emerges  and remains stable under further loading (Fig.\ref{fig:macro}E, Supplementary Movie S1). In this regime, we assume that the stress is evenly distributed among the bonds of this load-bearing subnetwork. These bonds have a density $n$ and a tension $\tau$, resulting in a macroscopic stress
\begin{equation} \label{eq:Sigma_affine}
    \sigma = n \tau
\end{equation}
In our model, we further assume that an increase $\delta \gamma$ of the macroscopic strain results in an equal stretch $\delta\epsilon=\delta\gamma$ of the load-bearing fibers, \emph{i.e.} that the system is differentially affine. This implies that the macroscopic differential modulus $K$ at large  stress directly reflects the microscopic differential stiffness $k=d\tau/d\epsilon$ of the load-bearing fibers:
\begin{equation} \label{eq:K_affine}
    K = n k
\end{equation}
Importantly, Eqs.~(\ref{eq:Sigma_affine}) and (\ref{eq:K_affine}) imply strong constraints connecting the stress- and density-dependence of the bulk modulus. Indeed,  for fibers with power-law stiffening, $k\propto \tau^x$, we find at large stress that 
\begin{equation}\label{eq:Gp}
    K \propto \sigma^x n^{1-x} 
\end{equation}
This equation implies that a single exponent $x$ of the fiber-level CL controls both the $n$- and $\sigma$-dependence of the macroscopic modulus. Strikingly, for collagen-like fibers with $x=1$, we find $K\propto \sigma$, independently of $n$. Our simple differentially affine model thus recapitulates the observations for collagen at large stress. While an alternate explanation involving normal stresses was previously proposed under shear~\cite{licup2015stress}, our model proposes a simple and general rationalization of the density-independence of the nonlinear elastic modulus of collagen.

Our differentially affine model (Eq.\eqref{eq:Gp}) also makes predictions for other CLs. For linear elements ($x=0$), we recover a stress-independent modulus  proportional to $n$. For $0<x<1$, $K(\sigma)$ increases with the load-bearing bond density. Remarkably, if $x>1$, we predict that, $K(\sigma)$ decreases with $n$. This startling behavior can be understood by considering the loading of a set of two bonds in parallel. If one of these segments is cut, the load is transferred to the remaining bond, doubling its load. If $x=1$,  however, this would also double the bond stiffness, thus leaving  the rigidity of the system unchanged.  For $x<1$, depleting the network leads to a reduced modulus. By contrast, if $x>1$ the stiffness of the remaining bond more than doubles, clarifying our counter-intuitive prediction that at constant stress, network depletion leads to stiffening.

We further confirm these predictions by simulating the fiber density- and stress-dependence of $K$ for networks with various fiber CLs (Fig.~\ref{fig:model}). Considering first the simple case of regular networks of variable mesh size (Fig.~\ref{fig:macro}G), where all fibers are load-bearing, we recover precisely the scaling behavior predicted by Eq.\eqref{eq:Gp} (Fig.~\ref{fig:macro}H-J) and, in particular, the $n$-dependence of the large-stress modulus (Fig.~\ref{fig:macro}H). For networks with random depletion (Fig.~\ref{fig:macro}E), the fiber density is controlled by the depletion parameter $p$, and the connection with the load-bearing fiber density $n$ is less evident.  Qualitatively however, we observe that the influence of $p$ is consistent with our prediction  (Fig.~\ref{fig:macro}F).  For all four CLs considered here, our model adequately captures the stress scaling of the differential modulus $K\propto \sigma^x$ (Fig.~\ref{fig:macro}A-D). 

The scaling of $K$ we observe can be compared  with macrorheology experiments that report a stiffening exponent $3/2$ for F-actin~\cite{gardel2004elastic}, fibrin~\cite{piechocka2010structural}, vimentin and neurofilaments networks~\cite{lin2010origins} and biomimetic hydrogels~\cite{jaspers2014ultra}, whereas Zn$^{2+}$-modified fibrin networks exhibit an exponent $1/2$~\cite{xia2021anomalous}. Our minimal model adequately captures the density dependencies observed for the different stiffening exponents (Fig.~\ref{fig:macro}H). In particular, for $x>1$, as for  WLC-fibers,  the differential modulus decreases for denser networks (Figs.~\ref{fig:macro}D,F), in agreement with earlier experiments~\cite{gardel2004elastic}. By contrast, for $0\leq x<1$ denser networks display an increased modulus (Figs.~\ref{fig:macro}C,F), as for the Zn$^{2+}$-modified fibrin networks~\cite{xia2021anomalous}. 
Taken together, our results show how at large stress the fiber density controls the mechanical response of soft heterogeneous networks in a way that depends sensitively on the nonlinear micromechanics of the constituents. Interestingly, collagen networks stand out by uniquely displaying a stress-controlled mechanical response  independently of network fiber density. 

\section*{Force ensures robust local response}\label{micro}

\begin{figure*}[ht]
  \centering
  \includegraphics[width = 1.0\textwidth]{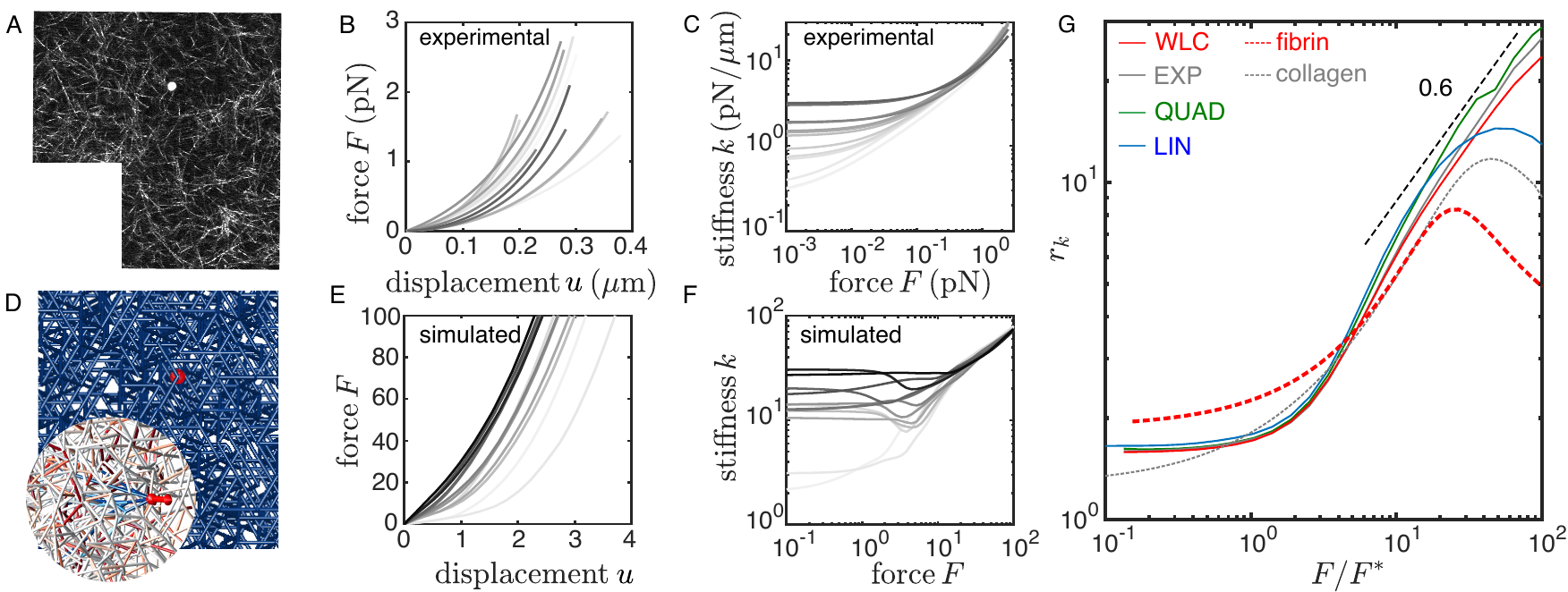}
  \caption{\textbf{Nonlinear response to local probing in disordered fibrous networks.} 
 Microrheology experiments on reconstituted collagen networks, A: confocal reflective image of a gel and a trapped 2 $\mathrm{\mu m}$ diameter bead (white dot), inset: schematic of the local probing using optical tweezers, B: applied force ($F$) versus measured displacement $u$ of the beads embedded at various locations, B: corresponding differential stiffness ($k$) versus $F$.
Local probing of 3D depleted EXP-fibers networks, D: a probe (red sphere) in the center of a  numerically generated disordered network, inset: tension distribution in fibers located behind the probe ($F=10$, same colorcode as in Fig.~\ref{fig:model}D).
Small random sample of measurements performed, E-F: same as B-C for the simulated probings.
G: Signal-to-noise ratio ($r_k$) of the stiffness measurements versus applied force, normalized by the force $F^\ast$ at the onset of the power law regime, for simulated networks with different CLs (Fig.~\ref{fig:model}) and microrheology measurements on collagen and fibrin gels. }
  \label{fig:micro}
\end{figure*}

\begin{figure*}[t]
  \centering
  \includegraphics[width=1.0\textwidth]{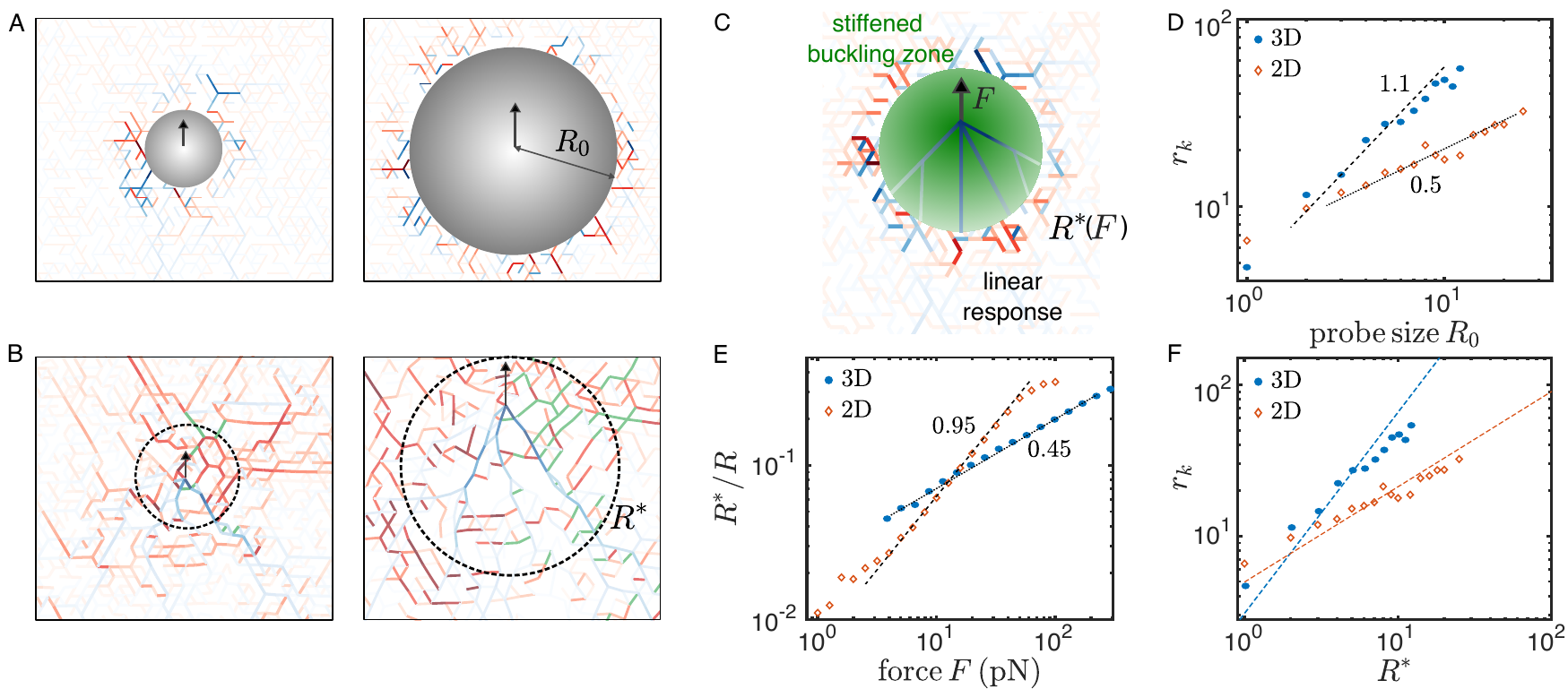}
  \caption{\textbf{Scale-dependent nonlinear mechanosensation.} Local probing of depleted EXP-fibers networks. 
  A: Tension distribution in 2D networks loaded by probes (gray circular objects) of two different radius $R_0$ ($F=0.001$, same tensions color code as in Fig.~\ref{fig:model}D). 
  B: Highlight of the nonlinear region $R^\ast$ (dashed circle) surrounding the point-force probe (arrow). The buckled fibers are shown in green (only on the right half of the figure).
  C: Schematics of the stiffened buckling zone embedded in a linearly deforming soft network.
  D: Signal-to-noise ratio ($r_k$) of stiffness measurements as a function of the probe radius ($R_0$) for an applied probe force $F=0.001$. 
  E:  Radius of the buckling region $R^\ast$, normalized by the system radius $R$, versus $F$ applied by a point-force probe.
  F: Comparison of the measured (dots and diamonds) and expected (dashed lines) signal-to-noise ratio, determined using Eqs.(\ref{eq:ratio_vs_F}) and (\ref{eq:Rs_vs_F}), as a function the size of the probe taken as $R^\ast$.  
  }
  \label{fig:EffectiveProbe}
\end{figure*}

To determine the macroscopic mechanical information a cell can obtain by performing local mechanical measurements inside a disordered network, we assume that cells are ideal mechanosensors: they probe the network by actively exerting a force at the scale of the network mesh-size and measure the network's local compliance without error. To conceptually understand how cells can locally perceive their mechanical environment in a simple way, we study the response of fiber networks to point-like force monopoles.

Experimentally, we perform active microrheology experiments on reconstituted collagen gels. We embed beads that are large enough to be trapped in the collagen mesh (Fig.~\ref{fig:micro}A). Using optical tweezers, we apply controlled forces on these beads to locally measure the mechanical response at various locations in the network (Fig.~\ref{fig:micro}A, see Materials and Methods and~\cite{Yang2022submitted}). This setup is  informative for mechanosensing as it allows us not only to probe the network at the cellular scale, but also to apply forces that are large enough to locally trigger the nonlinear response of a collagen matrix, as observed in the vicinity of cells embedded in fibrous matrices~\cite{han2018cell,van2016strain,hall2016fibrous}. The force-displacement curves obtained at various locations in the same network (Fig.~\ref{fig:micro}B) reveal two key features: i) the curves exhibit large variability, with a broad distribution of displacements at any force level, and ii) they are nonlinear and exhibit a pronounced stiffening response.  
For each force-displacement curve  $F(u)$ we measure locally, we determine the differential stiffness $k = {\rm d}F/{\rm d}u$ as a function of force (Fig.~\ref{fig:micro}C). 
Indeed, while it has been suggested that cells could be sensitive to several quantities such as the strain energy~\cite{panzetta2019cell},  viscoelastic  properties~\cite{chaudhuri2016hydrogels}, and stiffness with extensive evidence~\cite{engler2006matrix,guilak2009control,lo2000cell,isenberg2009vascular,chowdhury2010material,shellard2021durotaxis,hall2016fibrous}, a complete determination of the mechanical variables cells respond to is still lacking, especially in nonlinear environments. However, as cells have been shown to  adapt to the stiffness in collagen matrices, including the local differential stiffness increased by the forces exerted by the cell~\cite{hall2016fibrous}, we here characterize the cell-scale mechanical response in terms of this differential mechanical quantity.  

At low forces ($F<0.01 pN$), $k$ is independent of $F$ and this linear response is highly heterogeneous. Thus, at low forces cells can only acquire unreliable information about the mechanical landscape of their environment, as previously observed~\cite{grill2021directed,beroz2017physical,proestaki2019modulus,jones2015micromechanics,hayn2020inhomogeneities}. As the force is increased, however, the network stiffens, with locally softer networks stiffening at lower probe forces. Remarkably, at large forces ($F\geq 0.1pN$), the stiffness no longer strongly varies relative to the mean (Fig.~\ref{fig:micro}C).
In this nonlinear regime, the local stiffness that cells could measure inside a fibrous matrix becomes reliable: the probing force sets the measured stiffness, as observed for cells in collagen~\cite{han2018cell,hall2016fibrous}, and this measured stiffness becomes increasingly robust to local fluctuations in fiber density.

To theoretically understand this robustness of nonlinear stiffness measurements, we employ the model introduced in Fig.~\ref{fig:model}. We simulate local loading induced by a point-force monopole in a large spherical network with fixed boundary conditions (Fig.~\ref{fig:micro}D, Materials and Methods). To avoid boundary effects and correlations between individual measurements, we perform a local mechanical probe in the center of independently sampled network configurations. For collagen-like fibers with EXP CL, we measure many statistically independent force-displacement curves (Fig.~\ref{fig:micro}E) and the resulting differential stiffness (Fig.~\ref{fig:micro}F). We find that these simulations exhibit the same general trend as we observed experimentally: the differential stiffness is highly fluctuating in the  linear response regime where soft bending deformations dominate the response~\cite{beroz2017physical}, but tends towards a single $k(F)$ curve for a large force.  Thus, both our experimental and simulated behaviors reveal that the local mechanical response of a  collagen matrix becomes progressively reliable and insensitive to inherent structural disorder in the network when the force triggers the system's nonlinear response. 

This may appear unsurprising in light of the fiber density-independence of the nonlinear macroscopic modulus of a collagen network at fixed stress~\cite{licup2015stress} (Fig.~\ref{fig:macro}A, Eq.\ref{eq:Gp}). However, to show that the observed macroscopic density-independence of collagen networks does not explain the convergence of the microscopic $k(F)$ curves, which become independent of local fiber density heterogeneity, we perform microrheological simulations on networks with other fiber CLs (Figs.\ref{fig:model}A-C) whose nonlinear bulk stiffness is \emph{not} independent of average fiber density. Surprisingly, we observe the same features for all CLs (Fig.S5): the local stiffness strongly fluctuates in the linear response regime, while the different differential measurements robustly tend to a single master curve at large forces.  

To further quantify this increased robustness, we characterize the ensemble of independent stiffness values at a given force in terms of the signal-to-noise ratio $r_k = \langle k\rangle/\mathrm{std}(k)$ (Fig.~\ref{fig:micro}G). As $r_k$ increases with force, the mechanical signal becomes stronger relative to the stiffness heterogeneity. Interestingly, we find that the growth of $r_k$ with increasing force can be approximated as a power law in the nonlinear regime:
\begin{equation}\label{eq:ratio_vs_F}
   r_k =  \frac{\langle k\rangle}{\mathrm{std}(k)} \sim F^\alpha
\end{equation}
We measure $\alpha \approx 0.6$ for both simulated and experimentally measured mechanical responses. This exponent is also independent of network dimensionality  (Fig.~S1). In addition, $\alpha$ does not vary substantially when changing the CL: convergence of measurements is observed in all cases according to this universal power law. 

We experimentally confirm this predicted constituent-independence of the exponent $\alpha$ with microrheology measurements that we perform on a reconstituted fibrin network (Fig.~S11), with nonlinear bulk properties that are distinct from collagen~\cite{jansen2013cells}. Indeed, in fibrin we observe a marked increase of $r_k$ in the nonlinear response regime, with a power law increase similar to that observed in collagen (Fig.~\ref{fig:micro}G).
Therefore, the origins of this microrheological robustness cannot be the same as for the macrorheological convergence of nonlinear bulk modulus, which is specific to collagen.

In summary, at low force a single mechanical measurement is a poor estimator of the network's average mechanical properties: mechanosensing is strongly limited by structural heterogeneities. By contrast, our experiments and simulations indicate that the local mechanical response of fiber networks becomes largely insensitive to structural disorder at large force.

\section*{Nonlinear mechanosensing model \label{probesize}}

As the increased robustness of local micromechanical measurements at large force is generically observed for a range of CLs, its physical origin must lie in the fibrous structure of the network, rather than in the specific micromechanical properties of its constituents. This hypothesis is further supported by the modest influence of the CL on differential stiffness in the nonlinear regime: while at the macroscopic scale the bulk modulus followed $K\sim \sigma^x$ at large stress, the exponent $x$ does not set the microscopic stiffening response~\cite{Yang2022submitted} (Fig. S5G). 
Building on recent theoretical results showing that the nonlinear response of fiber networks to force dipoles results in an effective increase of the dipole size~\cite{ronceray2016fiber}, we propose that this robustness can instead be understood in terms of an effective increase of the size of the probed region. As network heterogeneities average out on larger scales, this would imply that, as the force increases, the probe becomes less sensitive to local density fluctuations.

To explore this idea, we first examine the linear local response and investigate how the size of the probe affects relative stiffness fluctuations of both 2D and 3D networks with EXP-fibers. While a measurement integrates mechanical contributions at all scales, the rapid decay of linear elastic deformations~\cite{landau1986theory} leads to a dominant contribution of density fluctuations in a small volume in the vicinity of the probe~\cite{beroz2017physical}. Consequently, individual measurements strongly depend on their location and the corresponding stiffness values display a large variability. A larger probe, however, samples the local mechanics of a bigger region in the vicinity of the network, thereby averaging more effectively over structural heterogeneity. Larger probes are thus more informative about the system's macroscopic response.

To demonstrate that increasing probe size indeed leads to more robust measurements, we perform simulations of a circular rigid body of radius $R_0$ applying a small force monopole ($F = 0.001$) to the network. (Fig.~\ref{fig:EffectiveProbe}A). 
For each probe radius $R_0$, we compute the signal-to-noise ratio of linear stiffnesses, revealing a power-law increase 
\begin{equation}\label{eq:ratio_vs_R0}
    r_k  \sim R_0^{\beta}
\end{equation}
with $\beta \simeq 0.5$ in 2D and $\beta\simeq 1.1$ in 3D. This power law increase of $r_k$ with probe size confirms that the responses depend less on local fiber density fluctuations. 

We map this increased robustness of local sensing with probe size to the increased robustness we observe in the nonlinear response for large local probing forces (Fig.~\ref{fig:micro}G). Thus, we argue that a sufficiently large applied force triggers the response of the network over an effectively larger region than in the linear response regime. To determine the force-dependent length-scale that sets the local response, we note that at the onset of the nonlinear response, fibers start to buckle near the probe (Fig.~\ref{fig:EffectiveProbe}B, left). This buckling spreads to a larger region in the network as the applied force increases~\cite{ronceray2016fiber} (Fig.~\ref{fig:EffectiveProbe}B, right). This leads to the emergence of a `buckling zone' of growing size $R^\ast$ with a large density of buckled fibers and the formation of tensed rope-like structures.  
Consequently, within the buckling zone the network strain stiffens as the elastic response is dominated by stretching of the ropes, which is a much stiffer mode of deformation than the fiber bending modes that govern the linear response regime.  
A probe force deforms both the network inside the buckling zone and the surrounding network beyond $R^\ast$. These two network sections thus effectively act in series. Because the buckling zone strain stiffens, however, it becomes much stiffer than the network section beyond $R^\ast$ that is still dominated by soft bending modes (Fig.~\ref{fig:EffectiveProbe}C). To first approximation, the buckling zone therefore becomes effectively rigid, and the compliance in response to the probe is dominated by the linear response of the network beyond  $R^\ast$. 
Thus, elastic nonlinearity renders the network disorder irrelevant inside the stiffened buckling zone, and local stiffness fluctuations are instead determined  by network disorder outside the nonlinear zone. Put simply, the emergent length scale $R^\ast$ renormalizes the size of the local probe in a force-dependent way. Local probes with large enough forces thus effectively probe the local linear mechanical properties of the network over a larger length scale making the response less sensitive to local disorder.

To understand the force-dependence of the buckling zone radius induced by monopole probing, we perform an analysis similar to previous work on dipole-induced buckling~\cite{rosakis2015model,xu2015nonlinearities,ronceray2016fiber}. Away from the probe, the stress decays as  $\sigma (r) \sim F/r^{D-1}$ due to force conservation. From the buckling condition, here written in terms of stress as $\sigma \sim \sigma_b$, with $\sigma_b$ the buckling stress, we expect $\sigma (R^\ast) \sim \sigma_b$. Therefore,  buckling occurs over a region of size
\begin{equation}\label{eq:Rs_vs_F}
R^\ast \sim F^\zeta     
\end{equation}
with $\zeta = {1/(D-1)}$. Indeed, we measure $\zeta \simeq 0.45$ in 3D and $0.95$ in 2D in our simulations (Fig.~\ref{fig:EffectiveProbe}E, Figs.~S7\&S8). 

To complete our nonlinear mechanosensing model, we now quantitatively connect the increase of $r_k$ with probe size in the linear response regime (Eq.\eqref{eq:ratio_vs_R0}) to the power law increase of $r_k$ with the applied point force (Eq.\eqref{eq:ratio_vs_F}). To do so, we  identify the effective probe size $R_0$ induced by nonlinear effects with the buckling length-scale $R^\ast$. Using Eq.\eqref{eq:ratio_vs_F} \& \eqref{eq:Rs_vs_F}, our nonlinear mechanosensing model predicts a power law increase of $r_k$ with the buckling range $R^\ast$: $r_k \sim {R^\ast}^{ \alpha/\zeta}$. Importantly, for both 2D and 3D simulated responses the expected power law increase is consistent with Eq.\eqref{eq:ratio_vs_R0} (Fig.~\ref{fig:EffectiveProbe}F, Supplementary Table 1). Thus, our scaling model establishes that locally probing the network in the nonlinear regime can be conceptualized as a linear probe with a renormalized probing radius associated to the radius of the buckling zone. This model quantitatively explains the CL-independent increase of mechanosensation reliability at large forces in fiber networks.

\section*{Nonlinear mechanosensation is reliable}

\begin{figure}
\centering
\includegraphics[width=\linewidth]{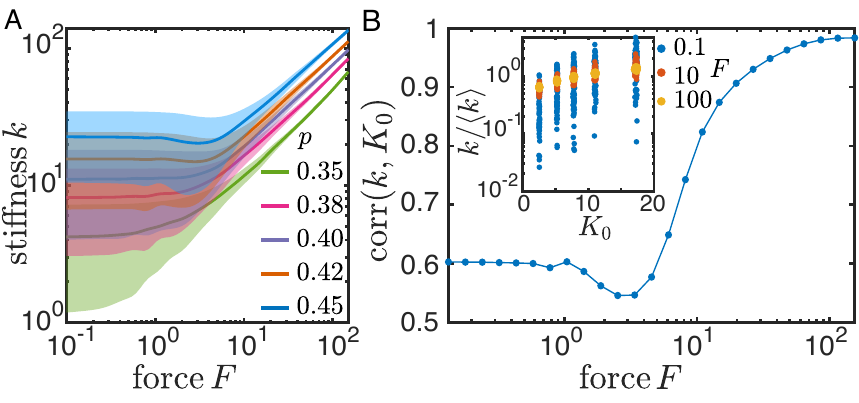}
\caption{\textbf{Robust nonlinear mechanosensing. }A: Mean stiffness as a function of force for different values of $p$, the shaded areas show the stiffness standard deviation. B: Pearson correlation coefficient between the local stiffness ($k$) and the linear bulk modulus ($K_0$) as a function of the local probing force. Inset: local stiffness measured at 3 different probing forces versus $K_0$. The normalization $\langle k \rangle$ is the mean of the measurements computed over all values $k$ measured at a given force level. }
\label{fig:mechanosensing}
\end{figure}

Can cells make use of nonlinear mechanosensation to reliably infer the large-scale mechanical properties of the surrounding matrix from local stiffness measurements? To address this question, we consider local stiffness measurements performed in networks with varying mean fiber density (Fig.~\ref{fig:mechanosensing}A), which sensitively tunes the network macroscopic modulus~\cite{licup2015stress,gardel2004elastic,piechocka2010structural,lin2010origins,jaspers2014ultra,xia2021anomalous,das2012redundancy,broedersz2011criticality}. We find that in the low force regime, there is a large overlap of the local stiffness measurements obtained on networks with different $p$ values: the variation of individual measurements exceeds the difference in mean stiffness of networks with different fiber densities. By contrast, at large forces the stiffness measured on networks with different fiber densities become clearly separated: by probing the mechanical response with large force, a cell can thus robustly infer the average stiffness of its surroundings and discriminate between the mechanical properties of networks of varying fiber density. 

To quantify the ability to infer the macroscopic properties of the network via nonlinear mechanosensing, we compare the local stiffness measured at different forces and the linear bulk modulus $K_0$ of various networks. While the average local stiffness increases with the bulk modulus, the large scatter of local measurements at low probing forces would prevent a reliable local estimation of the bulk stiffness (Fig.~\ref{fig:mechanosensing}B, inset). By contrast, at large forces, the relative mechanical heterogeneity of local measurements diminishes:  it becomes possible to discriminate networks with distinct bulk moduli using local stiffness measurements. Indeed, we find that the local stiffness and bulk modulus are increasingly correlated with increasing probe force (Fig.~\ref{fig:mechanosensing}B). The Pearson correlation coefficient approaches 1 at large forces: nonlinear mechanosensing enables accurate local inference of bulk network properties.

\section*{Discussion}

To understand the physical limits of nonlinear mechanosensing we studied the response of disordered networks to local force probes at the cell scale. Using a fiber network model and microrheology experiments, we demonstrated that a local probe can reliably determine both the local and macroscopic mechanical response of the disordered network by triggering either the linear response or elastic nonlinearities. This nonlinear mechanosensation becomes progressively insensitive to network disorder with increasing force. Nonlinear mechanosensing thus offers cells a reliable strategy to both locally determine and control the mechanical properties of a disordered ECM. 

We showed that macroscopic and microscopic nonlinear responses of disordered fiber networks are set by distinct mechanisms.
Macroscopically, the differential bulk stiffness is a power law  of the applied stress in the nonlinear regime (Figs.~\ref{fig:macro}A-D), with an  exponent $x$ set by the constitutive stiffening of a single fiber (Fig.\ref{fig:model}C). We  capture this macroscopic behavior with a differentially affine model. This model predicts that the load-bearing fiber density dependence of the differential bulk modulus  is set by an exponent $1-x$ (Eq.\eqref{eq:Gp}), which thus solely depends on the constitutive fiber stiffening (Fig.~\ref{fig:macro}H). Remarkably, this implies that the differential modulus of collagen ($x=1$) becomes insensitive to fiber density at large stress (Fig.~\ref{fig:macro}A), as observed experimentally~\cite{licup2015stress}. Our model further offers  insights into experiments with different stress and density-dependencies on various reconstituted networks with distinct stiffening behaviors~\cite{licup2015stress,gardel2004elastic,piechocka2010structural,lin2010origins,jaspers2014ultra,xia2021anomalous}. 
Microscopically, the local differential stiffness also stiffens with force as a power law (Fig. S5), but with an exponent that is not determined by the constitutive fiber stiffening~\cite{Yang2022submitted}. Instead, this local  stiffening is controlled by force-induced buckling and network nonlinearity in the form of a bending-stretching stiffening transition, giving rise to a stiffened buckling zone embedded in a linearly responding network (Fig.~\ref{fig:EffectiveProbe}C). Thus, local force-stiffening is caused by the effective probing of the linear network stiffness at increasingly larger scales set by the buckling zone radius. Our results on the difference between microscopic and macroscopic stiffening mechanisms could be used to further develop accurate approaches for 3D traction force inference or stress inference around cells in 3D matrices~\cite{han2018cell,hall2016fibrous}. 

The emergence of the stiffened buckling zone explains the   mechanical robustness to local force probes. A sufficiently large force effectively probes the linear response averaged over the structural disorder of an enlarged region of the network, with dimension set by the  force-dependent buckling length-scale.  This elastic regime is relevant for cell-ECM mechanical interactions. 
Indeed, several cell types apply traction forces of the order of tens of nanonewtons~\cite{trichet2012evidence,freyman2002fibroblast,tan2003cells,legant2010measurement,balaban2001force}. These forces can trigger the nonlinear response of reconstituted networks, as shown by our microrheology measurements in collagen and fibrin  (Fig.~\ref{fig:micro}B, Figs.S11). This nonlinear response is consistent with observations of contractile cells in a matrix: fibers buckle~\cite{han2018cell} and displacements are enhanced, decreasing more slowly than predicted by linear elasticity~\cite{hall2016fibrous,winer2009non,notbohm2015microbuckling}. 
Furthermore, the network stiffens in the wake of a cell applying traction~\cite{han2018cell,hall2016fibrous,winer2009non}, and our nonlinear mechanosening model can be used to determine the stiffness such cells would perceive. The magnitude of the force cells need to exert on their substrate to employ nonlinear mechanosensation \emph{in vivo} likely depends on context, where softer environments typically require smaller forces to trigger nonlinearity. Thus, even though neurons apply smaller forces~\cite{athamneh2015quantifying} than \emph{e.g.}, fibroblasts or cardiomyocites~\cite{balaban2001force}, differences in stiffness and nonlinear force thresholds of the natural surroundings could still allow such different cells to employ nonlinear mechanosensation. 

The formation of the force-controlled buckling zone is qualitatively independent of the local structure of the probe forces. In particular, the characteristic length-scale emerging in response to dipole loading is well characterized~\cite{ronceray2016fiber,xu2015nonlinearities,sander2013alignment}. We thus expect a similar robustness increase as a power law for dipole mechanosensors, but with modified exponents. 
Since cells are mechanically better described as force dipoles~\cite{schwarz2013physics}, this anticipated power law increase in response to dipole loading further supports nonlinear mechanosensing as a cellular strategy. 

Many aspects of cellular mechanosensation are still debated, including the internal cellular machinery and processes that are involved, as well as the mechanical variable that can be sensed by cells~\cite{saez2005mechanical,yip2013cellular}. Here, we characterized the reliability of the mechanical response in terms of stiffness, a mechanical property that is experimentally shown to influence cellular behavior~\cite{engler2006matrix,guilak2009control,lo2000cell,isenberg2009vascular,chowdhury2010material,shellard2021durotaxis,hall2016fibrous}. Yet, cells are also found to respond to other mechanical variables, such as the substrate strain energy~\cite{panzetta2019cell}. Therefore, one could consider other mechanical quantities to assess the limits of nonlinear mechanosensing, such as strain and elastic energies, which we characterized as alternatives (Fig.~S6). In these cases, the signal-to-noise ratio also generically increase with force for networks with a range of fiber constituents. This is understandable since local sensing becomes nonlocal due to the emergence of the buckling zone that facilitates disorder averaging over an increased length scale, regardless of the precise mechanical variable that is considered. 
Therefore, we argue that enhanced nonlinear mechanosensing is a general characteristic of disordered fibrous networks.

Finally, our work suggests that cells could employ nonlinear mechanosensation as a strategy to reliably sense and respond to the stiffness of their environment. In the linear regime, cells face a highly heterogeneous mechanical landscape~\cite{grill2021directed,beroz2017physical,proestaki2019modulus,jones2015micromechanics,hayn2020inhomogeneities}. If this linear  regime dominated cell-ECM interactions, we would expect the mechanical heterogeneity perceived by cells to lead to erratic stiffness-dependent cell behaviors following the large local stiffness fluctuations in the ECM. In contrast, if cells trigger the nonlinear response with large forces, then our nonlinear mechanosensation model  implies that cells instead face a strikingly homogeneous mechanical landscape that is directly correlated with the matrix' macroscopic modulus, and where stiffness-dependent cell behavior would not depend randomly on the cell's location in the matrix. Nonlinear mechanosensation also has  implications for the cell's ability to control the stiffness of their environment. The idea that cells actively stiffen their matrix and adapt in response to the enhanced stiffness has long been introduced~\cite{winer2009non,hall2016fibrous,jansen2013cells}, and it was speculated that such a feedback mechanism aims at reaching a specific substrate resistance. We here propose that nonlinear mechanosensation allows cells to exploit this mechanical feedback to accurately control their surrounding stiffness despite the inherent randomness of their local environment, allowing them to robustly perform mechanosensitive cellular functions even in a highly disordered ECM.

\section*{Materials and Methods}

\subsection*{Random network generation} Networks are generated by placing straight fibers on an ordered triangular (2D) or Face Centered Cubic (3D) lattice. These networks are randomly depleted with a bond occupation probability $p$. Unless stated otherwise, we use $p = 0.6$ in 2D and $p=0.4$ in 3D. 

\subsection*{Mechanical model} Fibers are discretized with bonds of rest length $\ell_0 = 1$. Each bond acts as a spring with linear stretching modulus $\mu = 100$ and their nonlinear longitudinal response $f(\epsilon_{ij})$ is described by a CL displayed in Fig.~\ref{fig:model}.  Fibers also resist transverse deflection with a bending rigidity $\kappa = 1$ that penalizes deflections of angle $\theta$. Fibers are connect by freely deforming hinges at their intersection. We consider a probe located on sites $i$ of position $\bold{R}_i$ and applying a force $\bold{F}_i$. The Hamiltonian of the system is
\begin{equation*}
    \mathcal{H} = - \sum_{\mathrm{forces}\, i} \mathbf{F}_i\cdot \mathbf{R}_i +  \sum_{\mathrm{bonds}\,\langle i,j\rangle} f(\epsilon_{ij}) + \sum_{\mathrm{hinges}\,\langle i,j,k\rangle}2\sin^2\frac{\theta_{ijk}}{{2}}
\end{equation*}

\subsection*{Macroscopic loading} The boundaries of the network are displaced to impose isotropic dilatation. Our depleted networks have dimensions $30\times 30 \times 30$ and periodic boundary conditions are imposed. The results are averaged over 3 independent random networks. The bulk stress and bulk modulus are computed from the first and second derivative of the system's energy: 
\begin{equation*}
    \sigma = \frac{1}{V}\frac{\partial \mathcal{H}}{\partial \gamma},\, K = \frac{1}{V}\frac{\partial^2 \mathcal{H}}{\partial \gamma^2}
\end{equation*}
where $\gamma$ is the applied dilatation strain and $V$ the system's volume.

\subsection*{Local probing} A point force $F$ is applied to a vertex at the center of spherical (circular in 2D) depleted networks of radius $R = 40$ in 3D ($R=100$ in 2D) for QUAD-, EXP- and WLC-fibers, and $R = 45$ for LIN-fibers. Finite size effects on the stiffness statistics are displayed in Fig. S3. The probe loading direction is $[0,1,0]$, not following any fiber direction in the undeformed network. The network boundaries are fixed. For each CL, 100 disorder realizations are performed.
To determine $r_k$ vs. $R_0$ (Fig.~\ref{fig:EffectiveProbe}D), the same boundary conditions apply, a sphere (disk) of radius $R_0$ is placed in the system's center ($R=30$ in 3D, refer to Fig. S4 for a characterization of finite size effects and $R=200$ in 2D) and applies a force $F = 10^{-3}$ in the direction $[0,1,0]$ to all incident bonds, 100 networks realizations are used.

\subsection*{Numerical resolution} 
At each applied loading, mechanical equilibrium is obtained by minimizing the total energy using the GNU Scientific Library BFGS implementation of the Broyden-Fletcher- Goldfarb-Shanno algorithm.

\subsection*{Microrheology
experiments} 
Microparticles (2 $\mathrm{\mu m}$ in diameter, C37278, ThermoFisher) are embedded in 4 $\mathrm{mg/mL}$ collagen gel and 3 $\mathrm{mg/mL}$ fibrin gel. A homemade optical tweezer is used to drag the particle at a velocity of 1 $\mathrm{\mu m/s}$. The displacements of the particles and the optical forces are recorded. Details of the experiments can be found in~\cite{Yang2022submitted}).

\textbf{Acknowledgements:} E.B. and C.P.B.: This project has received funding from the European Union’s Horizon 2020 research and innovation programme under the Marie Sklodowska-Curie grant agreement No 891217 and the Deutsche Forschungsgemeinschaft (DFG, German
Research Foundation) - Project-ID 201269156 - SFB 1032
(Project B12). P.R. is supported by France 2030, the French National Research Agency (ANR-16-CONV-0001) and the Excellence Initiative of Aix-Marseille University - A*MIDEX. M.G. and H.Y. acknowledge support from NIH 1R01G140108.\\
\textbf{Competing interests:} The authors have no competing interests to declare.

\bibliography{mechanosense}

\renewcommand{\thefigure}{S\arabic{figure}}
\setcounter{figure}{0}

\section*{Supplementary Information}

\section*{Numerical response}
\subsection*{Local probing of 2D networks}

\begin{figure*}[h!]
\centering
\includegraphics[width=0.8\textwidth]{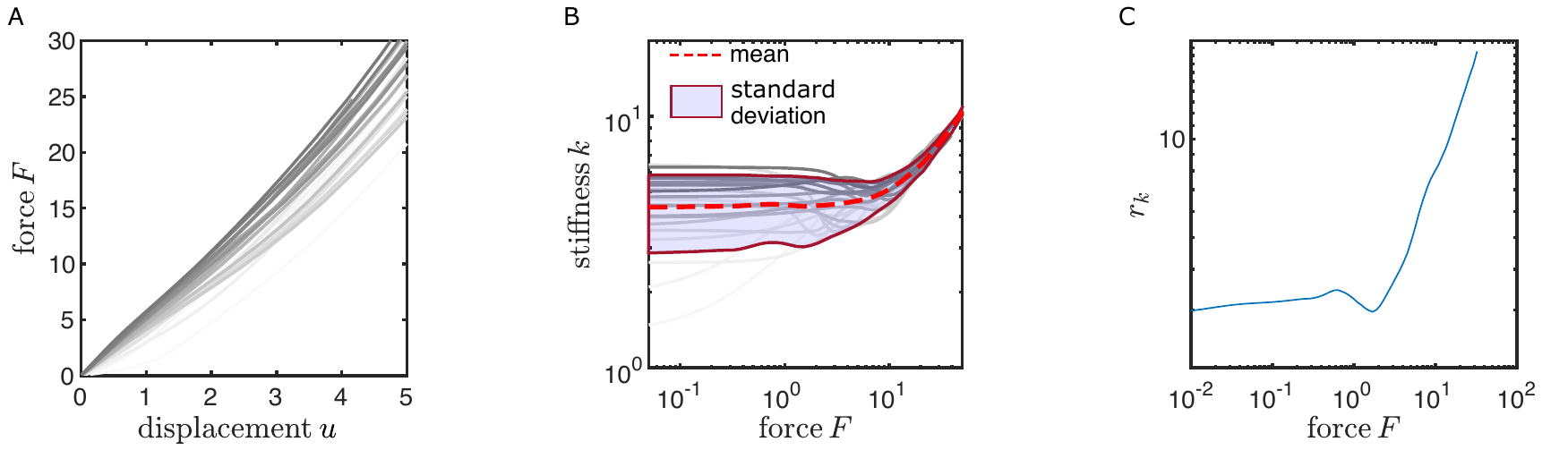}
\caption{\textbf{Statistics of local stiffness measurements of 2D depleted EXP-fibers networks.} Small random sample of measured A: force-displacement responses, B: corresponding stiffness versus force, C: associated signal-to-noise ratio as a function of the probing force.  } \label{fig:2D}
\end{figure*}

We also perform local differential stiffness measurements on 2D depleted networks of EXP-fibers. As for 3D networks, the force-displacement responses display both a strong variability and a nonlinear behavior (Fig.~\ref{fig:2D}A). Also consistent with 3D networks behavior, the corresponding differential stiffness measurements show a clear change of behavior as the force is increased: Individual measurements strongly fluctuate at low forces while they tend towards a master curve in the nonlinear regime (Fig.~\ref{fig:2D}B). We find that the increased reliability of the measurements in the nonlinear regime is also robustly obtained for these networks, with the signal-to-noise ratio following the same characteristic power law increase generically obtained for 3D networks (Fig.~\ref{fig:2D}C).

\begin{figure*}
\centering
\includegraphics[width=0.8\textwidth]{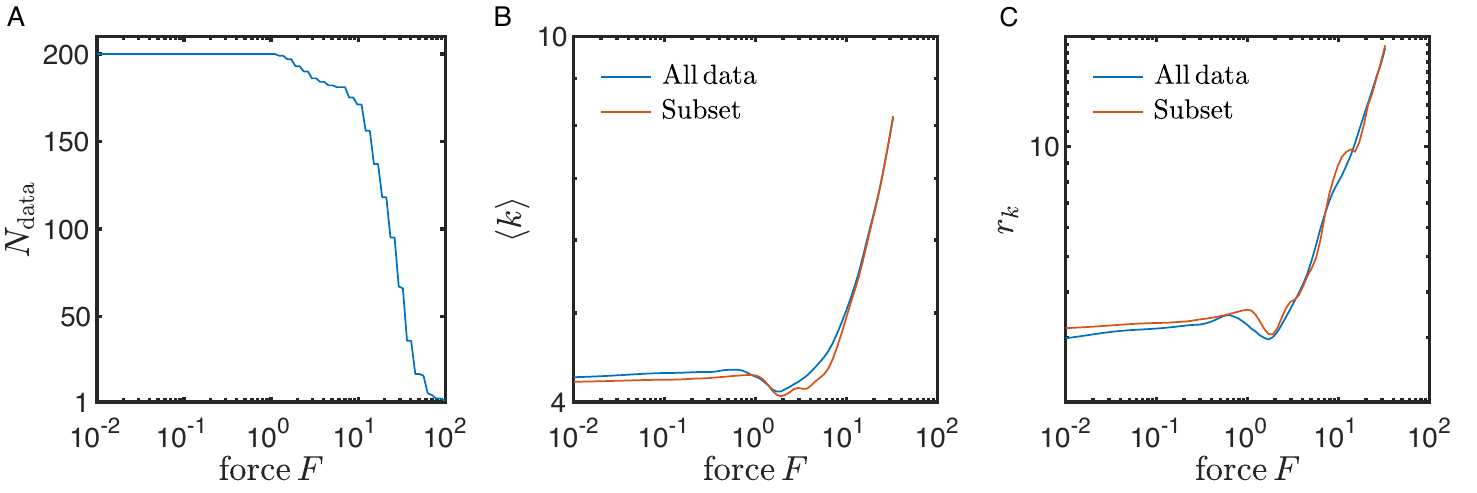}
\caption{\textbf{Sampling of local stiffness measurements of 2D depleted EXP-fibers networks.} A: Number of converged data points as a function of the applied force. Comparison of A: the mean stiffness and B: the signal-to-noise ratio versus force when all data points are considered at a given force (blue curves) or only the subset of simulations that are converged at all forces (orange curve).  } \label{fig:Subset}
\end{figure*}

We only consider simulations for which we confirm numerical convergence. The signal-to-noise ratio is thus computed over a decreasing number of data points as the applied force increases (Fig.~\ref{fig:Subset}A). We verify that the statistics of the stiffness ensemble remains unchanged if only the a subset of individual measurements is considered. In Fig.~\ref{fig:2D}C, $r_k$ is computed at a given force $F$ only if at least 40 data points are available. Considering the subset of stiffness measurements that are converged from the lowest to the highest force, Fig.~\ref{fig:Subset} shows that the obtained mean stiffness  and signal-to-noise ratio as a function of force is similar to that obtained in Fig. 2F of the main text over an ensemble of varying size.

\subsection*{Finite size effect: point-force probing} 

\begin{figure*}
\centering
\includegraphics[width=0.8\textwidth]{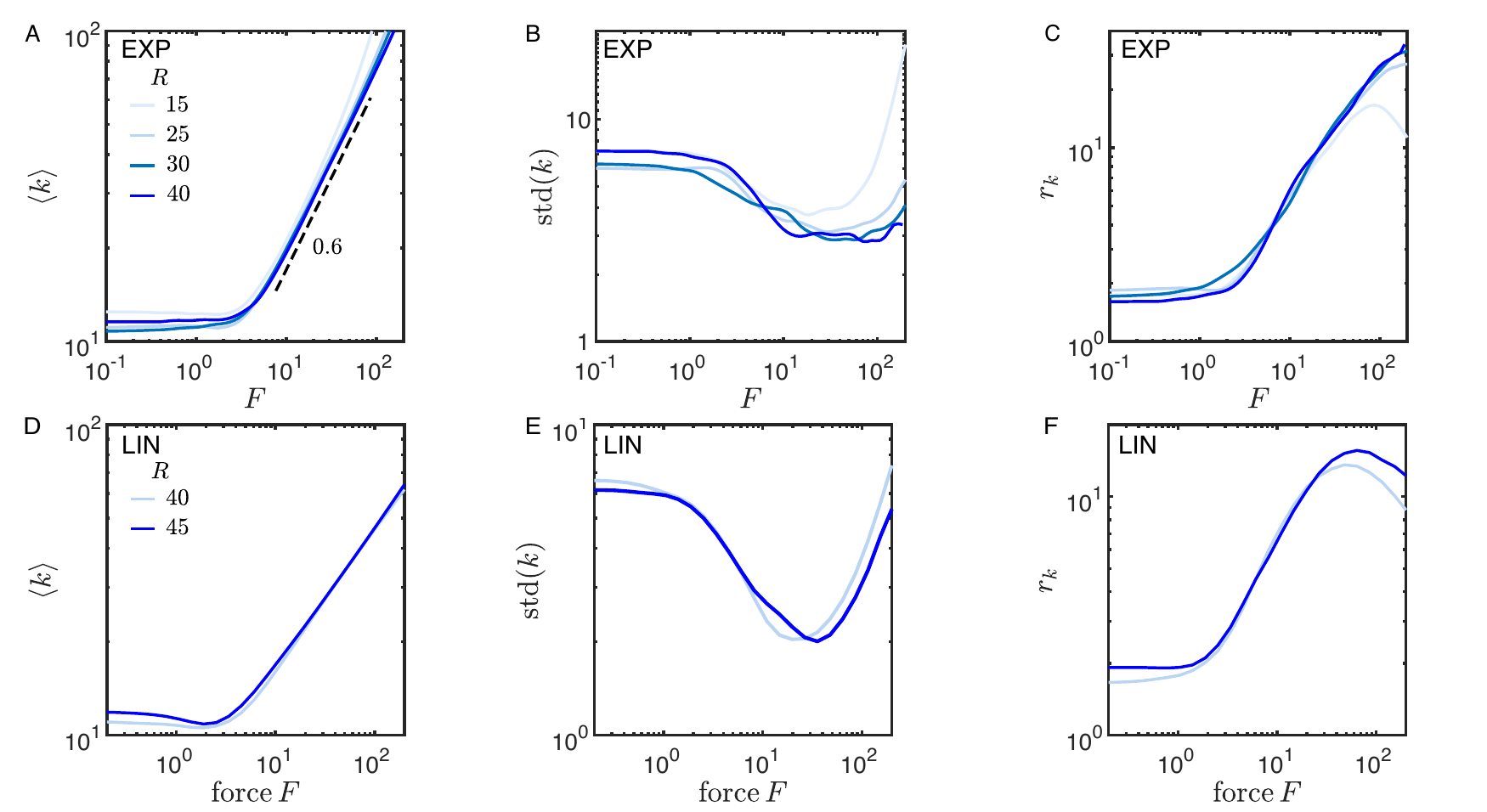}
\caption{\textbf{Finite size effect during point-force probing.} Mechanical response of 3D depleted EXP-fiber networks of various radius $R$, A: mean stiffness, B: stiffness fluctuations, C: signal-to-noise ratio as a function of the applied force $F$. D-F: same as A-C for LIN-fiber networks.} \label{fig:FSE}
\end{figure*}

Networks for our simulation are generated on a spherical domain of radius $R$. We verify the effect of the finite system size of 3D EXP-fibers networks on the statistics of the stiffness measurements (Fig.~\ref{fig:FSE}). At large forces, the network radius affects the average local response (Fig.~\ref{fig:FSE}A). In addition, we observe a sharp increase of stiffness fluctuations that is $R-$dependent at these large forces (Fig.~\ref{fig:FSE}B). These two contributions result in a saturation of the signal-to-noise ratio at forces that vary with system size (Fig.~\ref{fig:FSE}C). The radius $R = 40$ used in the main text ensures a nonlinear response over a large force range  before finite size effects set the mechanical response. We note that finite size effects occur at lower forces for fibers responding linearly in tension (LIN-fibers, Fig.~\ref{fig:FSE}D-F). We thus use $R=45$ for these networks.

\subsection*{Finite size effect: varying probe sizes}
\begin{figure*}
\centering
\includegraphics[width=0.3\textwidth]{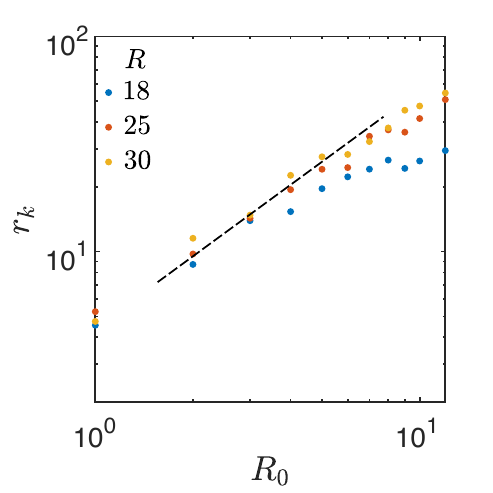}
\caption{\textbf{Finite system size effect during local probing}. Effect of the system radius $R$ of 3D depleted EXP-fibers networks on the local stiffness signal-to-noise ratio as a function of the probe size $R_0$. Model parameters are as described in Materials and Methods.} \label{fig:FSE_R0}
\end{figure*}

We determine the probe size radius at which the finite size of the network affects the statistics of local stiffness values in our simulations. Fig.~\ref{fig:FSE_R0} reveals that as the probe size $R_0$ exceeds an $R-$dependent value, the signal-to-noise ratio increases more slowly with the probe size. Therefore, the estimate of the $\beta$ exponent in Eq.(5) is performed on the largest system of radius $R = 30$ and for $R_0\leq 8$.

\subsection*{Local probing of networks of different fiber constituents}

\begin{figure*}
\centering
\includegraphics[width=\textwidth]{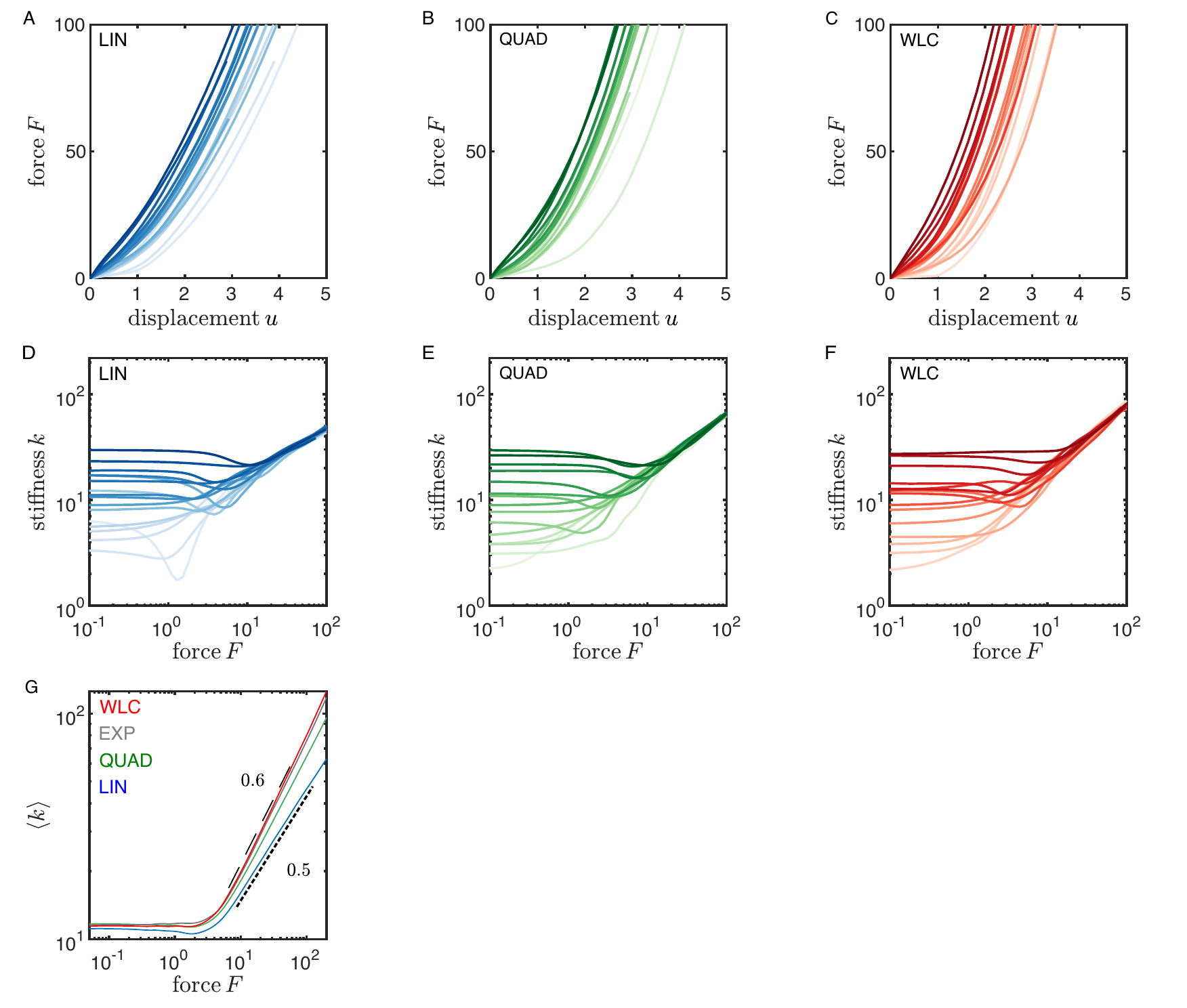}
\caption{\textbf{Robust nonlinear mechanical response of networks of various constituents.} Effect of the fibers constitutive law on the local mechanical response of 3D depleted networks. Force-displacement curves for a small sample of measurements performed for A: LIN-, B: QUAD- and C: WLC-fibers. D-F: corresponding differential stiffness versus force. G: Mean of the local stiffness measurements as a function of the force for the different CLs considered.} \label{fig:CL}
\end{figure*}

To evaluate the role of fiber-level nonlinearity in limiting nonlinear mechanosensation, we study the mechanical response to local point-force probing of networks constituted of fibers with a range of stiffening behavior (Fig.~1). For all the CLs considered, the force-displacement individual measurements display a strong variability and a highly nonlinear behavior (Fig.~\ref{fig:CL}A-C). The corresponding local stiffness also behaves as previously observed for 3D EXP-fibers network: there are strong fluctuations in the linear regime and while the local stiffness converge towards a single curve at large force (Fig.~\ref{fig:CL}D-F). The mean stiffness as a function of the applied force reveals a mild effect of the fiber-level stiffening exponent $x$: LIN-fiber networks ($x=0$) follow $\langle k\rangle \sim F^{1/2}$, as expected from the stiff body of radius $R^\ast \sim F^{1/2}$ probing the linear response of the surrounding network~\cite{Yang2022submitted}. For other constituents with $x>0$, the mean stiffness increases slightly faster as $x$ increases, revealing a contribution of mechanical response of the stiff nonlinear region being probed.

\section*{Alternative mechanical characterization: nonlinear robustness of the energy}

As cells could be responsive to several mechanical variable determined by probing the mechanical response of the ECM,  we also evaluate the robustness of the mechanical response in terms of strain energy $W$, defined as the area under the force-displacement curve, as an alternative to local stiffness measurements. Considering the force dependency of the local stiffness:

\begin{equation}\label{eq:k}
 k(F)\sim \begin{cases}
    k_0, & \text{if $F<<F^\ast$} \\
    F^\alpha & \text{if $F>>F^\ast$},
  \end{cases}
\end{equation}

\noindent the strain energy scales with force as

\begin{equation}
 W(F)\sim\begin{cases}
    F^2 & \text{if $F<<F^\ast$}\\
    F^{3/2} & \text{if $F>>F^\ast$}.
  \end{cases}
\end{equation}

\begin{figure*}
\centering
\includegraphics[width=0.9\textwidth]{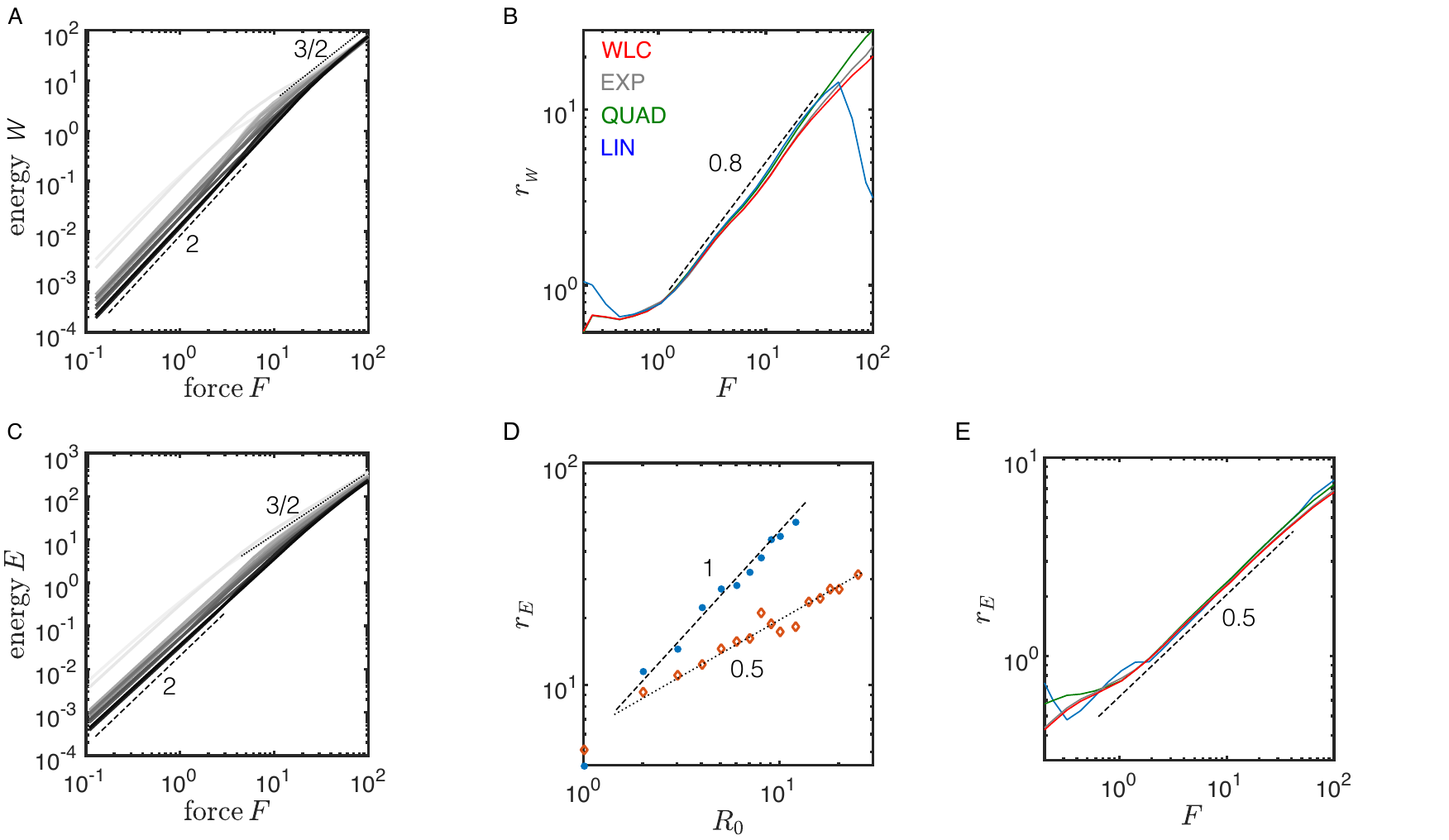}
\caption{\textbf{Mechanosensing via energy measurements}. A: Small sample of strain elastic energy $W$ versus applied force, measured in 3D depleted EXP-fiber networks. B: Signal-to-noise ratio of the strain energy measurements as a function of $F$ for $3D$ depleted networks of fibers with different CLs (Fig.1A-C). C: Same as A for the elastic energy $E$. D: Elastic energy signal-to-noise ratio $r_E$ as a function of the probe size for depleted EXP-fibers networks, the applied force is $F=0.001$. E: Same as B for the elastic energy $E$. } \label{fig:energy}
\end{figure*}

\noindent This force dependence of the strain energy is well recovered numerically for 3D depleted networks of EXP-fibers (Fig.~\ref{fig:energy}A). This sample of strain energy measurements as a function of the applied force again highlights the strong heterogeneity of local mechanical measurements and their non-linearity, as observed by characterizing the mechanical response in terms of stiffness.
For EXP-fibers network, we measure $r_E \sim F^{\alpha_W}$ with $\alpha_W \simeq 0.8$ (Fig.~\ref{fig:energy}B). Most importantly here, the power law exponent is independent of the fiber CL considered, consistent with our interpretation of the nonlinear mechanical response being set by the response at large scales. 

If one discards the loading history and focuses on the elastic energy $E = F\times u$, then using Eq.\eqref{eq:k}, the elastic energy also scales with $F$ as 

\begin{equation}
 E(F)\sim\begin{cases}
    F^2 & \text{if $F<<F^\ast$}\\
    F^{3/2} & \text{if $F>>F^\ast$},
  \end{cases}
\end{equation}

\noindent which is well recovered numerically (Fig.~\ref{fig:energy}C). 
According to our effective probe size interpretation, we can use the scaling $R^\ast \sim F^{1/2}$ and the measured signal-to-noise ratio dependency on the probe size $r_E \sim R_0^{1}$ (Fig.~\ref{fig:energy}D) to predict $r_E\sim F^{1/2}$. This power law scaling is confirmed numerically (Fig.~\ref{fig:energy}E).

\section*{Buckling zone characterization}

\subsection*{Determination from the compressive bond force}
\begin{figure*}
\centering
\includegraphics[width=0.5\textwidth]{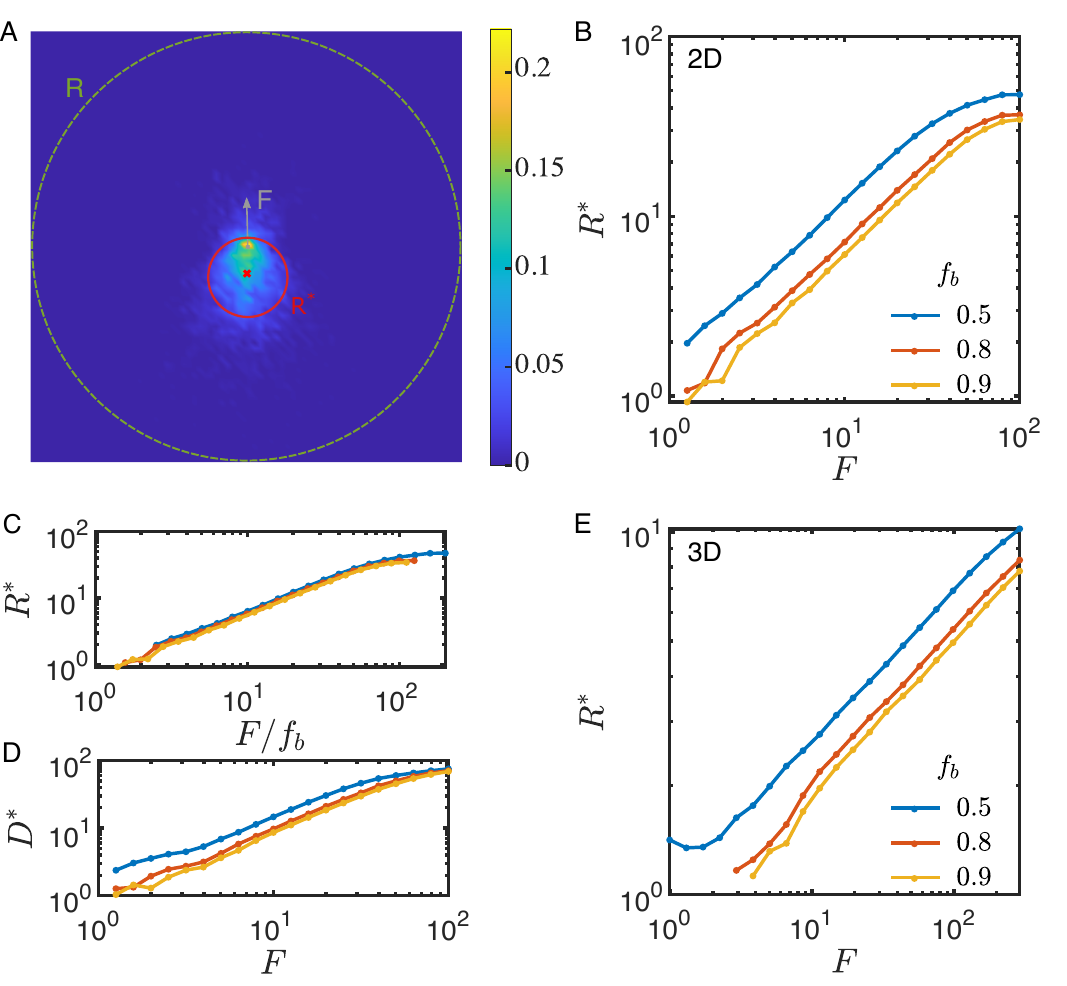}
\caption{\textbf{Determination of $R^\ast$ from the density of buckled bonds.} A: Density of buckled bonds of depleted 2D EXP-fibers networks, averaged over networks realizations, at a force $F=10$. The  red cross indicates the center of mass  of the density, the probe location and direction is indicated by the gray arrow,  the radius of gyration of the density, identified as $R^\ast$, is designated by the red circle and the boundary of the system by the dashed green circle. 
B: Power-law increase of $R^\ast$ with force for different buckling force thresholds $f_b$. C: Same as B with the horizontal axis rescaled by $f_b$. D: Spatial extent of the buckling region $D^\ast$ in terms of distance from the probe. E: $R^\ast$ versus $F$ determined on depleted 3D EXP-fibers networks of radius $R=25$ for different threshold values $f_b$.} \label{fig:Rs}
\end{figure*}

To determine the spatial extent of the buckling region, characterized by a high proportion of buckled fibers, we compute the radius of gyration of the density of buckled bonds. To do so, for a given disorder realization and a given applied force, we divide the network into squares (cubes) of size $\delta = 1.5$ and compute the density of buckled bonds as $\rho = N_b/N_T$, where $N_b$ is the number of buckled bonds and $N_T$ the number of bonds enclosed in that volume. At a given force, the density is averaged over simulations, leading to the example density map shown for 2D networks on Fig.~\ref{fig:Rs}A. We then successively determine the center of mass (red cross) and radius of gyration (red circle) of the density and identify the radius of gyration as a measure of $R^\ast$.

The constitutive law of the fibers accounts for bond-level buckling of the fibers via a softening response in compression (Fig. 1 of the main text). We set the state of the bond to be buckled if the compressive force on that bond is larger than a threshold force $f_b$. The value of threshold does not affect the exponent of the power law increase of $R^\ast$ with $F$ (Fig.~\ref{fig:Rs}B-C,E).  As displayed in  Fig.~\ref{fig:Rs}A, the buckling region is delocalized to the back of the probe loading (gray arrow) as a result of the lack of compressive forces transmission ahead of the probe. The nonlinear region thus extends over a distance $D^\ast = R^\ast + y_{\mathrm{CM}}$ from the probe, where $c_y$ is the location of the center of mass. This length scale approaches the radius $R$ of the system at large forces (Fig.~\ref{fig:Rs}D).

\subsection*{Determination from fiber reorientation}
\begin{figure*}
\centering
\includegraphics[width=0.9\textwidth]{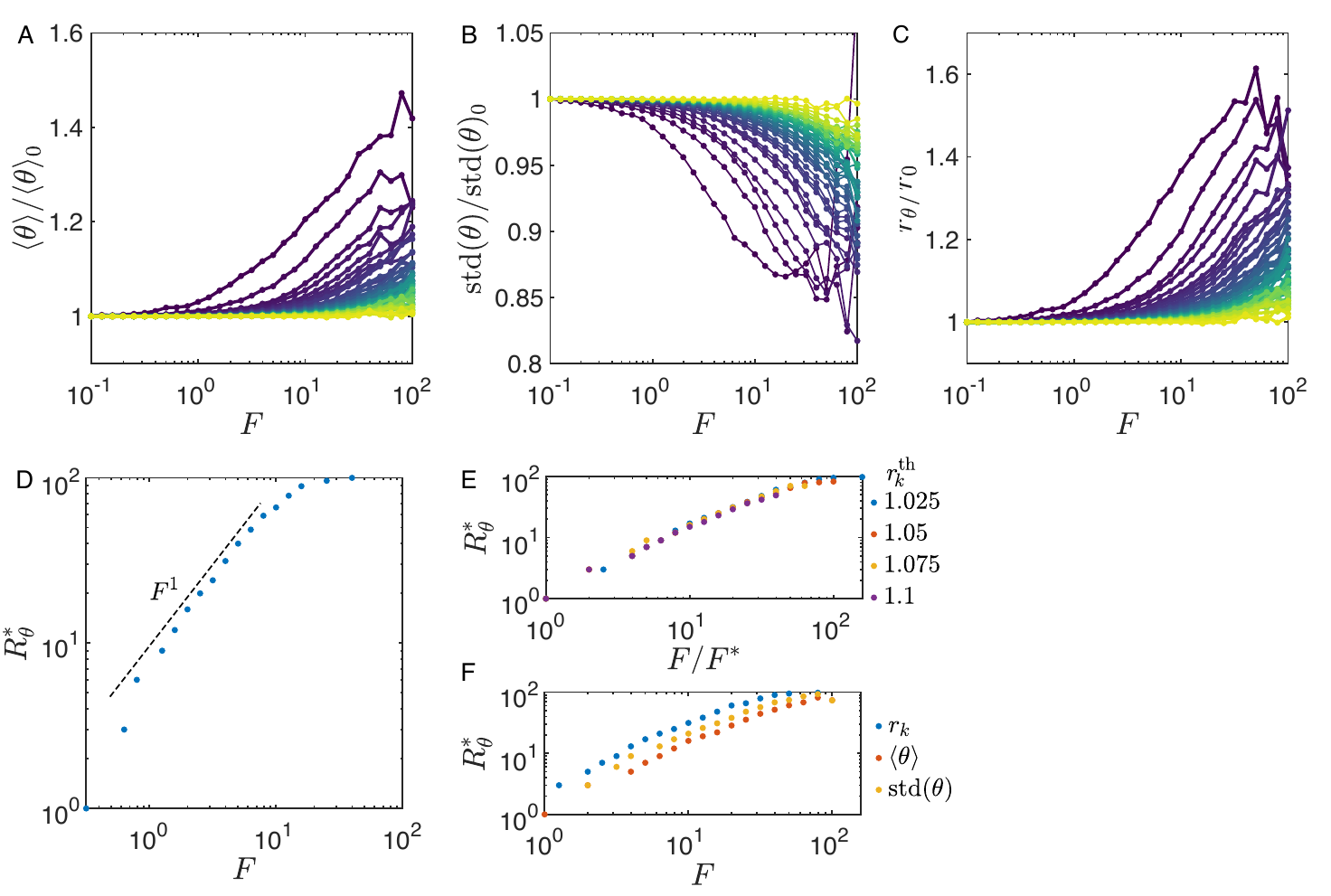}
\caption{\textbf{Determination of $R^\ast$ from bonds reorientation in 2D EXP-fiber networks}. Considering all bonds  located at a distance $[r,r+dr[$ from the probe, where $r$ varies from $0$ (dark blue) to $R-2$ (bright yellow) in increments of $dr=2$, A: mean, B: standard deviation and C: signal-to-noise ratio of bonds orientation as a function of the applied force. At each distance, these three quantities are normalized by their value at the lowest applied force. D: buckling region radius versus applied force, where $R^\ast$ is defined as the largest distance from the probe at which $r_k$ deviates by more than  $r_k^{\mathrm{th}} = 2.5\%$ from $r_0$. E: The value $r_k^{\mathrm{th}}$ considered does not affect the power law increase of the nonlinear region with force. F: $R^\ast$ can alternatively be determined from the variation of the mean bond orientation or fluctuations from their low force value.  } \label{fig:Rs_reorientation}
\end{figure*}

The loss of compressive resistance of buckled bonds is accompanied with large bond reorientations and the formation of the rope-like structure over $R^\ast$. Thus, the buckling region corresponds to a spatial domain where the initial geometry of the network is progressively replaced by a highly reorganized network (Figs. 4B). We use this observation to provide an alternative measurement of $R^\ast$ for 2D networks, although the same method can be extended to 3D systems. The statistics of bonds orientation at a given distance from the probe is shown as a function of force in Fig.~\ref{fig:Rs_reorientation}A-C. At each distance from the probe (low to high indicated by dark to bright curve), we determine the force at which the statistics deviates by more than a threshold amount from its value in the linear regime. Considering the deviation of the signal-to-noise ratio of the bond orientation $r_\theta$ by 2.5$\%$, we measure a  power law $R^\ast \sim F^1$  (Fig.~\ref{fig:Rs_reorientation}D) in good agreement with that determined from the density of buckled bonds. Again, the threshold selected does not affect the power-law exponent (Fig.~\ref{fig:Rs_reorientation}E) and one can perform this measurement on any of the three quantities $r_\theta$, $\langle \theta\rangle$ and $\mathrm{std}(\theta)$, as shown in Fig.~\ref{fig:Rs_reorientation}E.

\section*{Exponents of the scaling arguments}

The exponents $\alpha$, $\beta$ and $\zeta$ measured for depleted 2D and 3D EXP-fiber networks are summarized in Table~\ref{tab:Exponents}, where the values  expected from our scaling arguments, leading to $\zeta = \alpha/\beta$, are also displayed. 

\begin{table}\centering
\caption{Measured exponent values} \label{tab:Exponents}
\begin{tabular}{c|cc|cc}
     Exponent & \multicolumn{2}{c|}{2D} & \multicolumn{2}{c}{3D}  \\
              & measured & expected & measured & expected \\
              \hline 
    $\alpha$ & $0.6 \pm 0.1$ & $0.48 \pm 0.2$ & $0.6 \pm 0.15$ & $0.5 \pm 0.11$ \\
    $\beta$ & $0.5 \pm 0.2$ & $0.63 \pm 0.15 $ & $1.1 \pm 0.2$ & $1.3 \pm 0.3$ \\
    $\zeta$ & $0.95 \pm 0.15$ & $1.2 \pm 0.5$ & $0.45 \pm 0.05$ & $0.55 \pm 0.17$ \\
\end{tabular}
\end{table}

\section*{Experimental measurements}

The microrheology measurements are performed on three distinct reconstituted collagen networks and one fibrin gel. 
\subsection*{Collagen networks}

We label each ensemble of measurements performed on collagen as Set 1 to 3, where Set 3 is reproduced from~\cite{han2018cell}. The measurements performed in each network are shown in Fig.~\ref{fig:Outliers}
and the signal-to-noise ratio evolution with force of the different sets are compared in Fig.~\ref{fig:Comparison}.

\subsubsection*{Outliers removal}
\begin{figure*}
\centering
\includegraphics[width=0.9\textwidth]{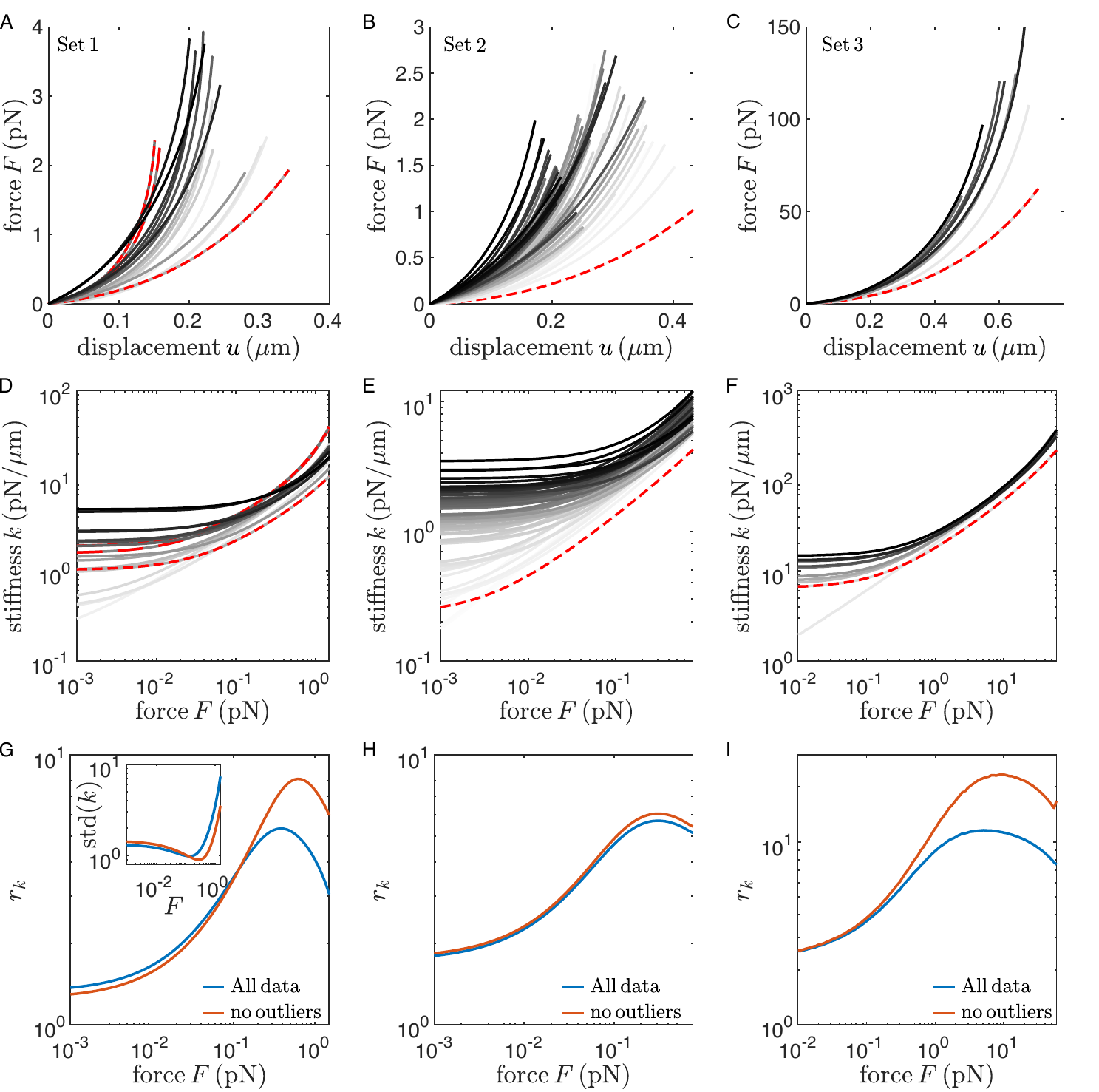}
\caption{\textbf{Statistics of stiffness measurements in reconstituted collagen networks.} For each network, A-C: force-displacement response and D-F: corresponding differential stiffness versus force applied by the beads. The red dashed lines indicate identified outliers. G-I: signal-to-noise ratio of the stiffness ensembles when considering all curves (blue) or discarding outliers (orange).} \label{fig:Outliers}
\end{figure*}

For each set, both the force-displacement responses (Fig.~\ref{fig:Outliers}A-C) and the corresponding differential stiffness measurements (Fig.~\ref{fig:Outliers}D-F) greatly vary from one location to another. For each data set, one to a few curves stand out and are identified as outliers (red dashed curves). These behaviors can be the result of several experimental limitations. Indeed, it is challenging to control the vertical location of the bead in the network. Yet, a bead located anomalously close to a boundary will lead to a finite size effect that results in an increased stiffening at low forces. The effect of removing these outlying curves when performing the computation of $r_k$ is shown on Fig.~\ref{fig:Comparison}G-I. Discarding these curves does not substantially affect the exponent of the power law increase of $r_k$ but ensures that it takes place over an extended force range. Thus, in the main text and in the next section we do not consider these outlying curves in computing the ensembles statistics.

\subsubsection*{Replicability}
\begin{figure*}
\centering
\includegraphics[width=0.35\textwidth]{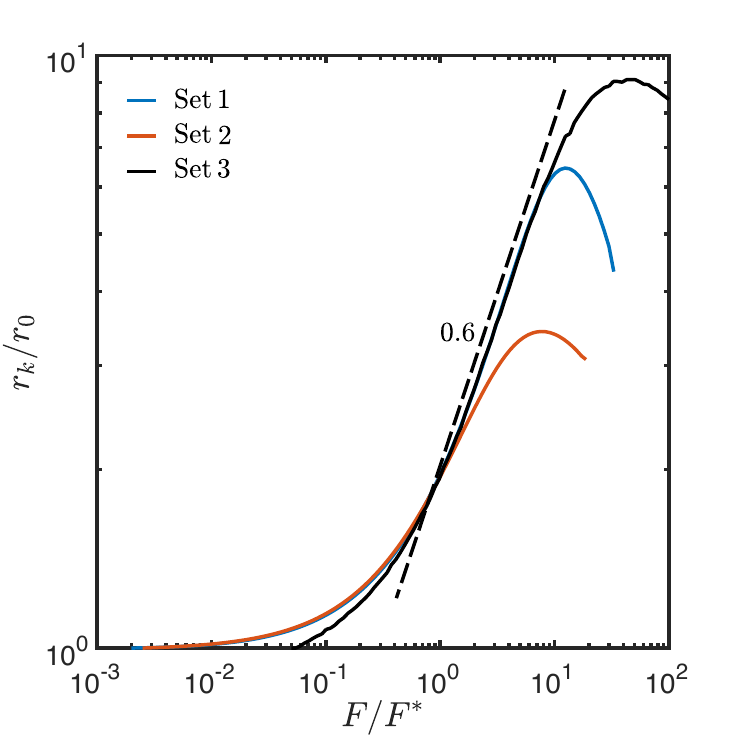}
\caption{\textbf{Replicability of microrheology measurements on reconstituted collagen networks}. Comparison of the ensemble of stiffness measurements obtained on three distinct reconstituted collagen networks. For each set of measurement, the normalized signal-to-noise ratio $r_k$ is shown as a function of the rescaled force $F$. Here, $r_0$ is the lowest value of $r_k$ and $F^\ast$ corresponds to the force onset of the power law response. } \label{fig:Comparison}
\end{figure*}

To compare the measurements performed on the distinct networks, we show the signal-to-noise ratio $r_k$ as a function of force in Fig.~\ref{fig:Comparison}. The curves are normalized by the lowest value $r_0$ and horizontal collapse of the curves is obtained by rescaling the force. The overlapping of the curves  reveals the  replicability of the experiments and strongly supports our theoretical interpretation.

\subsection*{Fibrin networks}

\begin{figure*}
\centering
\includegraphics[width=0.9\textwidth]{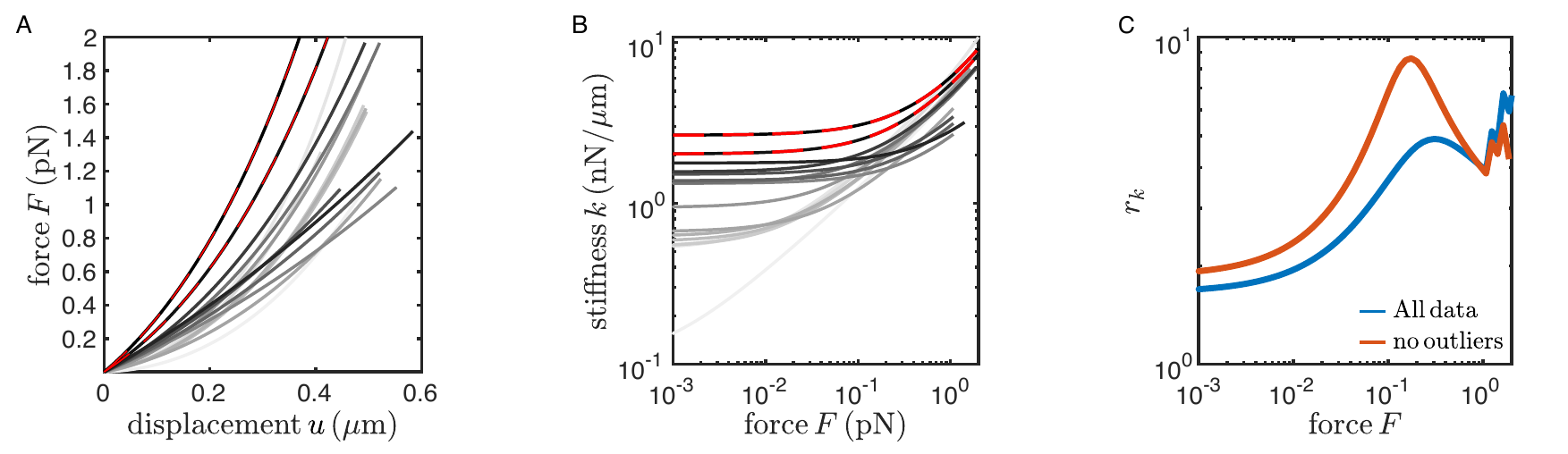}
\caption{\textbf{Statistics of stiffness measurements in a reconstituted fibrin network.} For each measurement, A: force-displacement response, B: corresponding differential stiffness versus force applied by the beads. The dashed red lines marks the identified outliers. C: Signal-to-noise ratio versus force.} \label{fig:fibrin}
\end{figure*}

The mechanical response of microrheology measurements on fibrin gels are displayed in Fig.~\ref{fig:fibrin}. We recover the characteristic variability and nonlinearity observed for local measurements in collagen networks and simulated loadings.

\section*{Movie description}

Our differentially affine model relies on the hypothesis that at asymptotically large stresses a tense subnetwork that carries most of the stress emerges  and remains stable under further loading. The emergence and stability of this subnetwork at large stresses is visible in Movie S1 which displays the ension distribution in a 2D depleted EXP-fiber network in response to macroscopic dilatation strain and the  corresponding mechanical response.

\end{document}